		\documentclass[10pt,journal,compsoc]{IEEEtran}
		\ifCLASSOPTIONcompsoc
		  \usepackage[no compress]{cite}
		\else
		  \usepackage{cite}
		\fi
		\ifCLASSINFOpdf
		\else
		\fi
		\usepackage{mathtools}
		\usepackage{stmaryrd}
		\usepackage{amssymb}
		\usepackage{amsmath}
		\usepackage{moreverb}
		\usepackage{amssymb}
		\usepackage{lineno}
		\usepackage{graphicx}
		\usepackage{amssymb}
		\usepackage{amsmath}
		\usepackage{moreverb}
		\usepackage{amssymb}
		\usepackage{lineno}
		\usepackage{graphicx}
		\graphicspath{ {images/} }
		\usepackage{caption}
		\usepackage{subcaption}
		\usepackage{algpseudocode}
		\usepackage{multirow}
		\usepackage[utf8]{inputenc}
		\usepackage[english]{babel}
		\usepackage{mathtools}
		
		\usepackage[colorlinks,bookmarksopen,bookmarksnumbered,citecolor=red,urlcolor=red]{hyperref}
		\usepackage{rotating,booktabs,comment,array,tabu,pdflscape}
		\usepackage{longtable,slashbox,color,colortbl}
		\usepackage{float}
		\usepackage{stmaryrd}
		\usepackage{multirow}
		\usepackage{epstopdf}
		\usepackage{makecell}
		\usepackage{multirow}
		\usepackage{booktabs}
		\usepackage{graphicx}
		\usepackage[table,xcdraw]{xcolor}
		\definecolor{Gray}{gray}{0.9}
		\usepackage[linesnumbered,boxed,lined, ruled]{algorithm2e}
		\let\oldnl\nl
		\newcommand{\nonl}{\renewcommand{\nl}{\let\nl\oldnl}}
		\usepackage{enumitem}
		\usepackage{tabularx}
		\usepackage{ltablex} 
		\usepackage{longtable}
		\usepackage{supertabular}
		\usepackage{color, colortbl}
		\usepackage{csquotes}
		\usepackage{subfloat}
		\usepackage[bottom]{footmisc}
		
		\usepackage{mathrsfs}
		
		\usepackage{stackengine}
		
		\usepackage{amsmath}

\begin{document}
			\fontsize{9.5pt}{11.5pt} 
			%
			\title{$\mu$-DDRL: A QoS-Aware Distributed Deep Reinforcement Learning Technique for Service Offloading in Fog computing Environments}

		\author{Mohammad~Goudarzi, 
			Maria~A.~Rodriguez, 
            Majid Sarvi, 
			and~Rajkumar~Buyya 
			\IEEEcompsocitemizethanks{\IEEEcompsocthanksitem 
				M. Goudarzi and R. Buyya are with the Cloud Computing and Distributed Systems (CLOUDS) Laboratory, School of Computing and Information Systems, The University of Melbourne, Australia.
				
				\IEEEcompsocthanksitem
				M. A. Rodriguez is with the School of Computing and Information Systems, University of Melbourne, Australia.

                \IEEEcompsocthanksitem
				M. Sarvi is with the Department of Infrastructure Engineering, University of Melbourne, Australia \protect\\
				E-mail: m.goudarzi@unsw.edu.au, marodriguez@unimelb.edu.au, majid.sarvi@unimelb.edu.au, rbuyya@unimelb.edu.au. 
				 
			}
		}
			
			%
			%

			\markboth{Submitted to Transactions Journal, VOL.YY, NO.ZZ, YEAR}%
			{Shell \MakeLowercase{\textit{et al.}}: Bare Demo of IEEEtran.cls for Computer Society Journals}
			%

			\IEEEtitleabstractindextext{
\begin{abstract}
Fog and Edge computing extend cloud services to the proximity of end users, allowing many Internet of Things (IoT) use cases, particularly latency-critical applications. Smart devices, such as traffic and surveillance cameras, often do not have sufficient resources to process computation-intensive and latency-critical services. Hence, the constituent parts of services can be offloaded to nearby Edge/Fog resources for processing and storage. However, making offloading decisions for complex services in highly stochastic and dynamic environments is an important, yet difficult task. Recently, Deep Reinforcement Learning (DRL) has been used in many complex service offloading problems; however, existing techniques are most suitable for centralized environments, and their convergence to the best-suitable solutions is slow. In addition, constituent parts of services often have predefined data dependencies and quality of service constraints, which further intensify the complexity of service offloading. To solve these issues, we propose a distributed DRL technique following the actor-critic architecture based on Asynchronous Proximal Policy Optimization (APPO) to achieve efficient and diverse distributed experience trajectory generation. Also, we employ PPO clipping and V-trace techniques for off-policy correction for faster convergence to the most suitable service offloading solutions. The results obtained demonstrate that our technique converges quickly, offers high scalability and adaptability, and outperforms its counterparts by improving the execution time of heterogeneous services.
\end{abstract}

	\begin{IEEEkeywords}
		Fog/Edge Computing, Internet of Things (IoT), Deep Reinforcement Learning (DRL), QoS-Aware Service Offloading. 
\end{IEEEkeywords}}

\maketitle

\IEEEdisplaynontitleabstractindextext

%
\IEEEpeerreviewmaketitle

\IEEEraisesectionheading{\section{Introduction}\label{sec:introduction}}
\IEEEPARstart{D}{ue} to continuous advancements of hardware and software technologies, Internet of Things (IoT) devices (e.g., sensors, cameras) have been widely adapted to many application scenarios, resulting in more intelligent and efficient solutions. The domain of IoT applications is diverse and ranges from smart transportation and mobility systems to healthcare care \cite{luo2021minimizing,qi2019knowledge}. To illustrate, ongoing urbanization along with ever increasing traffic has led to a high demand for smart transportation and mobility systems, where the main goal is to obtain faster, cheaper, and safer transportation by collecting, augmenting, and analyzing heterogeneous data generated from various sources. Smart transportation and mobility systems comprise a wide range of services with heterogeneous characteristics, including but not limited to vehicle automation, dynamic traffic light control, road condition monitoring, and vehicle localization and monitoring \cite{pekar2020application,liu2022cloud}. These growing IoT services, either computationally intensive or latency-critical, require high computing, storage, and communication resources for smooth and precise execution \cite{goudarzi2021distributedDDRL,tian2021user}. On their own, IoT devices with limited computing and storage capacities cannot efficiently process and analyze the vast amount of generated data in a timely manner, and hence they require surrogate resources for processing and storage. We refer to the process of allocating services on surrogate resources as offloading/outsourcing \cite{wang2019delay,goudarzi2020application}.
\par
\begin{figure}[!t]
	\centering 
	\includegraphics[width=\linewidth, height=4cm]{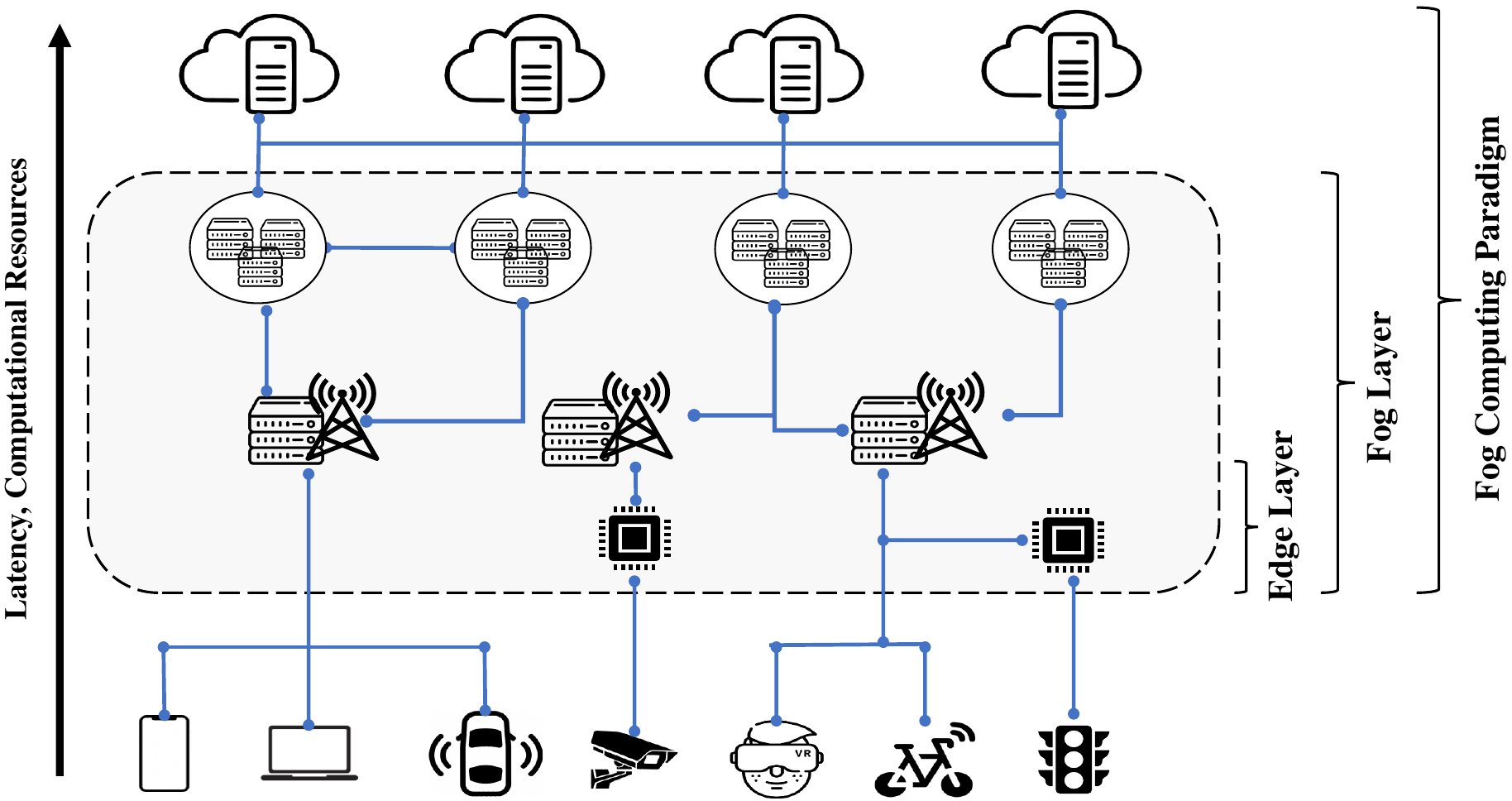}
	\caption{An overview of Edge and Fog computing}
	\label{fig:edgeandfogcomputing}
\end{figure}

A key enabler for the practical deployment of IoT services is cloud computing, which offers unlimited elastic surrogate computing and storage resources for the execution of applications. Through this paradigm, a subset of constituent parts of services (e.g., tasks) can be placed and executed on remote Cloud Servers (CSs) \cite{liu2022cloud, deng2021fogbus2}. Although the cloud computing paradigm suits many computation-intensive services, it cannot effectively satisfy the demands of latency-critical services, such as online traffic control systems and remote health monitoring systems, due to high communication latency and low bandwidth between IoT devices and CSs \cite{xu2019computation}. Fortunately, the fog computing paradigm offers a powerful infrastructure for carrying out complex latency-critical IoT services. In fog computing, distributed and heterogeneous Fog Servers (FSs), situated in the vicinity of IoT devices, provide computing and storage resources for IoT devices with lower access latency and higher bandwidth, properties that are critical for latency-sensitive services \cite{goudarzi2019fog,al2022ai}. However, FSs are resource-limited compared to CSs. Thus, they may not be solely capable of managing all sorts of services, specifically computation-intensive ones. In our view, Edge computing harnesses only the closest distributed resources in the vicinity of IoT devices, while fog computing harnesses FSs deployed in a hierarchical structure (containing Edge resources and other higher-level distributed resources) and cloud resources to address the demands of computation-intensive and latency-sensitive services (some works may interchangeably use these terms). Fig.~\ref{fig:edgeandfogcomputing} shows an overview of Edge and fog computing.
\par
Offloading services to fog computing environments enables the practical deployment of diverse computation-intensive and latency-critical services. However, because resources are limited compared to user requests, finding the offloading solution for the efficient execution of services is a difficult but significantly important task. Moreover, many services are often modeled as a Directed Acyclic Graph (DAG), in which vertices refer to tasks/microservices, while edges demonstrate data communication among dependent constituent parts \cite{goudarzi2021distributedDDRL,qi2019knowledge,al2017energy}. Such dependencies among the constituent parts of each service further increase the complexity of the service offloading problem. Finally, IoT services require a different level of Quality of Service (QoS) for smooth execution, usually considered as a set of constraints for the service offloading problem \cite{ding2021budget}. Considering the number of IoT services, surrogate resources, and applied constraints, offloading services to the fog computing environment is an NP-hard problem, meaning that finding the optimal solution in a timely manner is impossible as the problem space grows \cite{brogi2017qos,bittencourt2017mobility}. To solve this problem, many heuristics and rule-based algorithms have been proposed, which rely mainly on relevant knowledge abstracted by researchers, such as mathematical models. However, the fog computing environment is a stochastic and dynamic computing environment for which the above-mentioned class of algorithms is inefficient as they cannot adapt themselves to the ongoing changes in the environment \cite{qi2019knowledge}. 
\par
Because of the constant changes in the heterogeneous fog computing environment, service offloading decisions must be adaptively updated. Deep Reinforcement Learning (DRL) provides a promising solution by combining Reinforcement Learning (RL) with Deep Neural Network (DNN). The DRL agent can dynamically learn optimal policy and long-term rewards in stochastic environments without prior knowledge of the system using exploration \cite{wang2020fast}. However, the DRL agent requires collecting a large and diverse set of experience trajectories in the exploration phase (which is usually very time-consuming) that later will be used to train the agent to learn the optimal policy. Thus, centralized DRL techniques suffer significantly from high exploration costs and slow convergence, which negatively affect the performance of service offloading techniques in highly heterogeneous and stochastic fog computing environments. Moreover, most of the distributed DRL techniques employed for the service offloading problem do not use the experience trajectories generated by different actors efficiently, leading to slow convergence to optimal results.
\par
To solve the above-mentioned problems, we propose a distributed deep reinforcement learning technique, entitled $\mu$-DDRL, to efficiently offload heterogeneous services with different QoS requirements in heterogeneous fog computing environments. The $\mu$-DDRL uses Asynchronous Proximal Policy Optimization (APPO) \cite{petrenko2020sample}, which uses an actor-learner framework in which distributed agents are deployed on different FSs and work in parallel in the exploration phase. Thus, not only do the exploration costs significantly decrease (i.e., experience trajectories are generated faster by several agents), but also the experience trajectories are further diversified, resulting in more efficient learning of optimal service offloading policy. Furthermore, due to decoupled acting and learning, $\mu$-DDRL uses two important techniques, called V-trace \cite{espeholt2018impala} and PPO Clipping \cite{schulman2017proximal}, to cope with the policy gap between actors and the learner. The main contributions of this paper are summarized as follows.

%
			\begin{itemize}
				\item An execution time model for minimizing service offloading of DAG-based services while satisfying their required QoS is proposed.
				\item A distributed DRL-based service offloading technique, called $\mu$-DDRL, for dynamic and stochastic fog computing environments is proposed, which works based on the APPO framework. It helps to efficiently use the experience trajectories generated by several distributed actors to train a better service offloading model. Consequently, we propose a reward function for $\mu$-DDRL to minimize the execution time of each DAG-based service while satisfying QoS. We use two techniques, called V-trace and PPO Clipping, to solve the policy gap between actors and learners. 
				\item We conduct comprehensive experiments using a wide range of synthetic service DAGs derived from real-world services to cover the requirements of different services. Also, we compare the performance of $\mu$-DDRL with three state-of-the-art and widely adopted DRL and DDRL techniques.
			\end{itemize}
		  \par
			The rest of the paper is organized as follows. Section~\ref{relatedw} reviews related work on service offloading techniques in fog computing environments. Section~\ref{system} describes the system model and problem formulation. The DRL model and its main concepts are presented in Section~\ref{sec:DRLModel}. Our proposed DDRL-based service offloading framework is presented in Section~\ref{sec:placement}. In Section~\ref{sec:evaluation}, the performance of $\mu$-DDRL is studied and compared with state-of-the-art techniques. Finally, Section~\ref{conclusion} concludes the paper and draws future work.
\section{Related Work}
\label{relatedw}
In this section, we discuss service offloading techniques in fog computing environments.
\par
Several works in the literature consider an IoT service either as a single service or as a set of independent components. In these works, the constituent parts of each service can be executed without any data dependency among the different components. Bahreini et al.~\cite{bahreini2021vecman} proposed a greedy technique to manage the computing resources of connected electric vehicles to minimize energy consumption. Chen et al.~\cite{chen2019collaborative} proposed a collaborative service placement technique in an Edge computing environment, in which a network of small cell base stations is deployed to minimize the execution cost of services. The problem is first solved in a centralized manner to find the optimal solution, and then a graph coloring technique is proposed to solve the service placement problem in a decentralized way. Maleki et al.~\cite{maleki2021mobility} proposed location-aware offloading approaches based on approximation and greedy techniques to minimize the response time of applications in a stochastic and dynamic Edge computing environment. Badri et al.~\cite{badri2019energy} modeled the location-aware offloading problem in an Edge computing environment as a multistage stochastic problem, and proposed a Sample Average Approximation (SAA) algorithm to maximize the QoS of users while considering the energy budget of Edge servers. Ouyang et al.~\cite{ouyang2018follow} proposed a Markov approximation technique to obtain a near-optimal solution for the cost-performance trade-off of the service offloading in the Edge computing environment, which does not require prior knowledge of users' mobility. Zhang et al.~\cite{zhang2019task} and Huang et al.~\cite{huang2019deep} proposed a Deep Q-Network (DQN) based technique to learn the optimal migration policy and computation offloading in mobile Edge computing, respectively. Zhang et al.~\cite{zhang2019task} considered scenarios where user mobility is high, such as autonomous driving, to improve user QoS, while Huang et al.~\cite{huang2019deep} considered fixed IoT devices, such as smart cameras. Furthermore, Chen et al.~\cite{chen2018optimized} proposed the use of a more stable version of the DQN technique, Double DQN (DDQN), to learn the optimal offloading policy. Garaali et al.~\cite{garaali2022learning} proposed a multi-agent deep reinforcement learning solution based on the Asynchronous Advantage Actor Critic (A3C) to reduce system latency, so that each agent aims to learn interactively the best offloading policy independently of other agents. Li et al.~\cite{li2022deep} designed a hierarchical software-defined network architecture for vehicles that send their computational tasks to remote Edge servers for processing. Next, a DQN technique is proposed to solve the load balancing optimization problem so that the mean square deviation of the loads among different Edge servers is minimized. Liao et al.~\cite{liao2023online} developed a double RL-based offloading solution for mobile devices to minimize the energy consumption of devices and the delay of tasks based on the Deep Deterministic Policy Gradient (DDPG) and DQN, respectively. Ramezani et al.~\cite{ramezani2023task} proposed a Q-learning-based technique to reduce load imbalance and waste of resources in fog computing environments. Zhou et al.~\cite{zhou2023cost} proposed a Mixed Integer Non-Linear Programming (MINLP) formulation to jointly optimize the computation offloading and service caching in fog computing environments. To solve the optimization problem, an A3C-based technique is designed to optimize the offloading decision and service caching.
\par
Several works in the literature considered IoT services as a set of dependent tasks, modeled as DAG. In these services, each task can only be executed when its parent tasks finish their execution. Wang et al.~\cite{wang2019delay} proposed a mobility-aware microservice coordination scheme to reduce the service delay of real-time IoT applications in the fog computing environment. First, the authors formulated the problem as an MDP and then proposed a Q-learning technique to learn the optimal policy. The authors in~\cite{kimovski2021mobility} proposed a mobility-aware application placement technique for microservices based on the Non-Dominated Sorting Genetic Algorithm (NSGA-II) in a fog computing environment to minimize the execution time, energy consumption, and cost. The authors in~\cite{goudarzi2021distributed} proposed fast heuristics for the placement and migration of real-time IoT applications, consisting of microservices, in a hierarchical fog computing environment. Shekhar et al.~\cite{shekhar2019urmila} proposed a service placement technique for microservice-based IoT applications while considering a deterministic mobility pattern for users based on previous mobility data. The main goal of the centralized controller is to make the placement decision to satisfy the latency requirements. Qi et al.~\cite{qi2019knowledge} proposed an adaptive and knowledge-driven service offloading technique based on DRL for microservice-based smart vehicles' applications in an Edge computing environment. Mouradian et al.~\cite{mouradian2019application} and Sami et al.~\cite{sami2020vehicular} studied metaheuristic-based service placement in Edge and fog computing environments, respectively. Mouradian et al.~\cite{mouradian2019application} formulated the service placement problem as using Integer Linear Programming (ILP) and proposed a Tabu Search-based placement algorithm to minimize the response time and cost of applications. Sami et al.~\cite{sami2020vehicular} proposed an evolutionary Memetic Algorithm (MA) for the placement of containerized microservices in vehicle onboard units. Wang et al.~\cite{wang2020fast} proposed a centralized DRL technique based on Proximal Policy Optimization (PPO) for the placement of a diverse set of IoT applications to minimize their response time. Bansal et al.~\cite{bansal2022urbanenqosplace} proposed a dueling DQN technique to minimize the energy consumption of IoT devices and the response time of modular IoT applications in multi-access edge computing environments. 
\begin{table*}
\centering
\caption{A qualitative comparison of related works with ours}
\label{tab:relatedwork}
\resizebox{1\textwidth}{!}{%
\renewcommand{\arraystretch}{1.5}
\footnotesize
\begin{tabular}{|c|c|c|c|c|c|c|c|c|c|c|c|c|c|c|c|c|c|c|c|}
\hline
\multirow{4}{*}{Ref}                                   & \multicolumn{2}{c|}{Service Structure}                                                                  & \multicolumn{7}{c|}{Environmental Architecture}                                                                                                                  & \multicolumn{6}{c|}{Decision Engine}                                                                                                                                                                                                                                                                                                                                            \\ 
\cline{2-16}
& \multirow{3}{*}{Design} & \multirow{3}{*}{\begin{tabular}[c]{@{}c@{}}Granular\\Het\end{tabular}} & \multirow{3}{*}{Tiering} & \multicolumn{2}{c|}{IoT Device}      & \multicolumn{3}{c|}{Edge/Fog Servers}                            & \multirow{3}{*}{\begin{tabular}[c]{@{}c@{}}Multi\\Cloud\end{tabular}}  & \multicolumn{3}{c|}{Optimization Characteristics}                                                                                   & \multirow{3}{*}{\begin{tabular}[c]{@{}c@{}}Decision\\Planner\\Technique\end{tabular}} & \multirow{3}{*}{\begin{tabular}[c]{@{}c@{}}High\\Scalability\end{tabular}} & \multirow{3}{*}{\begin{tabular}[c]{@{}c@{}}High\\Adaptability\end{tabular}}  \\ 
\cline{5-9}\cline{11-13}
&                         &                                                                                  &                          & Number   & Het             & Number   & Het             & Cooperation               &                             & Perspective & \begin{tabular}[c]{@{}c@{}}Problem\\Modelling\end{tabular} & \begin{tabular}[c]{@{}c@{}}QoS\\Constraints\end{tabular} &                                                                               &                                                                            &                                                                              \\ 
\hline
\cite{bahreini2021vecman}                                                     & Monolithic              & \checkmark                                                        & Edge                     & Multiple & \checkmark & Multiple & \checkmark & $\times$    & $\times$      & IoT         & MINLP                                                      & $\times$                                   & Greedy                                                                        & $\times$                                                     & $\times$                                                       \\ 
\hline
\cite{chen2019collaborative}                                                     & Independent             & \checkmark                                                        & Edge                     & Multiple & ND                        & Multiple & ND                        & \checkmark & $\times$      & IoT         & IP                                                         & $\times$                                   & Optimal                                                                       & $\times$                                                     & $\times$                                                       \\ 
\hline
\cite{maleki2021mobility}                                                     & Monolithic              & $\times$                                                           & Edge                     & Multiple & \checkmark & Multiple & \checkmark & $\times$    & $\times$      & IoT         & MINLP                                                      & $\times$                                   & Approx                                                                        & $\times$                                                     & $\times$                                                       \\ 
\hline
\cite{wang2019delay}                                                    & Microservice            & \checkmark                                                        & Fog                      & Multiple & \checkmark & Multiple & \checkmark & \checkmark & $\times$      & IoT         & MDP                                                        & $\times$                                   & Q-learning                                                                    & $\times$                                                     & \checkmark                                                    \\ 
\hline
\cite{kimovski2021mobility}                                                     & Microservice            & \checkmark                                                        & Fog                      & Multiple & \checkmark & Multiple & \checkmark & \checkmark & $\times$      & Hybrid      & ND                                                         & $\times$                                   & NSGA2                                                                         & $\times$                                                     & $\times$                                                       \\ 
\hline
\cite{badri2019energy}                                                     & Monolithic              & $\times$                                                           & Edge                     & Multiple & \checkmark & Multiple & \checkmark & $\times$    & $\times$      & Hybrid      & IP                                                         & Energy                                                   & SAA                                                                           & $\times$                                                     & $\times$                                                       \\ 
\hline
\cite{goudarzi2021distributed}                                                     & Microservice            & \checkmark                                                        & Fog                      & Multiple & \checkmark & Multiple & \checkmark & \checkmark & $\times$      & IoT         & ILP                                                        & $\times$                                   & Heuristic                                                                     & $\times$                                                     & $\times$                                                       \\ 
\hline
\cite{ouyang2018follow}                                                    & Monolithic              & \checkmark                                                        & Edge                     & Multiple & \checkmark & Multiple & \checkmark & \checkmark & $\times$      & Hybrid      & Lyapu                                                   & Deadline                                                 & Approx                                                                        & $\times$                                                     & $\times$                                                       \\ 
\hline
\cite{shekhar2019urmila}                                                 & Microservice            & \checkmark                                                        & Fog                      & Single   & $\times$    & Multiple & \checkmark & $\times$    & $\times$      & IoT         & ND                                                         & $\times$                                   & GTB                                                                           & $\times$                                                     & $\times$                                                       \\ 
\hline
\cite{qi2019knowledge}                                                    & Microservice            & \checkmark                                                        & Edge                     & Multiple & \checkmark & Multiple & \checkmark & \checkmark & $\times$      & IoT         & MDP                                                        & $\times$                                   & A3C                                                                           & \checkmark                                                  & \checkmark                                                    \\ 
\hline
\cite{zhang2019task}                                                    & Monolithic              & $\times$                                                           & Edge                     & Multiple & $\times$    & Multiple & \checkmark & $\times$    & $\times$      & IoT         & ND                                                         & $\times$                                   & DQN                                                                           & $\times$                                                     & \checkmark                                                    \\ 
\hline
\cite{mouradian2019application}                                                     & Modular                 & \checkmark                                                        & Fog                      & Multiple & \checkmark & Multiple & \checkmark & ND                        & \checkmark   & IoT         & ILP                                                        & $\times$                                   & Tabu                                                                          & $\times$                                                     & $\times$                                                       \\ 
\hline
\cite{sami2020vehicular}                                                    & Microservice            & \checkmark                                                        & Edge                     & Multiple & \checkmark & Multiple & \checkmark & \checkmark & $\times$      & IoT         & ND                                                         & $\times$                                   & MA                                                                            & $\times$                                                     & $\times$                                                       \\ 
\hline
\cite{huang2019deep}                                                     & Monolithic              & ND                                                                                & Edge                     & Multiple & \checkmark & Single   & $\times$    & $\times$    & $\times$      & IoT         & MDP                                                        & $\times$                                   & DQN                                                                           & $\times$                                                     & \checkmark                                                    \\ 
\hline
\cite{chen2018optimized}                                                     & Independent             & \checkmark                                                        & Edge                     & Single   & \checkmark & Multiple & $\times$    & $\times$    & $\times$      & IoT         & MDP                                                        & $\times$                                   & DDQN                                                                          & $\times$                                                     & \checkmark                                                    \\ 
\hline
\cite{wang2020fast}                                                    & Modular                 & \checkmark                                                        & Edge                     & Multiple & \checkmark & Single   & $\times$    & $\times$    & $\times$      & IoT         & MDP                                                        & $\times$                                   & PPO                                                                           & $\times$                                                     & \checkmark                                                    \\ 
\hline
\cite{garaali2022learning}                                                    & Independent                 & \checkmark                                                        & Edge                     & Multiple & \checkmark & Multiple   & \checkmark   & $\times$    & $\times$      & IoT         & MDP                                                      & $\times$                                   & A3C                                                                           & \checkmark                                                    & \checkmark                                                    \\ 
\hline
\cite{li2022deep}                                                    & Independent                 & \checkmark                                                        & Edge                     & Multiple & \checkmark & Multiple   & \checkmark    & $\times$    & $\times$      & System         & MDP                                                        & $\times$                                   & DQN                                                                          & $\times$                                                     & \checkmark                                                    \\ 
\hline
\cite{liao2023online}                                                    & Independent                & \checkmark                                                        & Edge                     & Multiple & \checkmark & Single   & $\times$    & $\times$    & $\times$      & IoT         & MDP                                                        & $\times$                                   & DDPG-DQN                                                                           & $\times$                                                     & \checkmark                                                    \\ 
\hline
\cite{ramezani2023task}                                                    & independent                 & \checkmark                                                        & Fog                     & Multiple & $\times$ & Multiple   & ND    & $\times$    & $\times$      &    System      & MDP                                                      & $\times$                                   & Q-learning                                                                          & $\times$                                                     & $\times$                                                    \\ 
\hline
\cite{bansal2022urbanenqosplace}                                                    & Modular                 & \checkmark                                                        & Edge                     & Multiple & \checkmark & Multiple   & \checkmark    & \checkmark    & $\times$      & IoT         & MDP                                                         & $\times$                                   & DuDQN                                                                           & $\times$                                                     & \checkmark                                                    \\ 
\hline
\cite{zhou2023cost}                                                    & Independent                 & \checkmark                                                        & Fog                     & Multiple & \checkmark & Multiple   & \checkmark    & $\times$    & $\times$      & Hybrid         & MDP                                                        & $\times$                                   & A3C                                                                           & \checkmark                                                     & \checkmark                                                    \\ 
\hline
\begin{tabular}[c]{@{}c@{}}Our\\Technique\end{tabular} & Microservice            & \checkmark                                                        & Fog                      & Multiple & \checkmark & Multiple & \checkmark & \checkmark & \checkmark   & IoT         & MDP                                                        & Deadline                                                 & APPO                                                                          & \checkmark                                                  & \checkmark
\\
\hline
\multicolumn{16}{|l|}{\begin{tabular}[c]{@{}l@{}}Het: Heterogeneity, ND: Not Defined, MDP: Markov Decision Process, IP: Integer Programming, ILP: Integer Linear Programming, MINLP: Mixed Integer Non-Linear Programming, Lyapu: Lyapunov \\SAA: Sample Average Approximation, GTB: Gradient Tree Boosting, PPO: Proximal Policy Optimization, MA: Memetic Algorithm, DDQN: Double DQN, DuDQN: Dueling DQN \end{tabular}}
\\
\hline
\end{tabular}
}
\end{table*}
\par
The main elements of the related literature are identified in Table~\ref{tab:relatedwork}, consisting of service structure, environmental architecture, and decision engine. Also, this table compares our proposed work with the literature on the identified elements to further highlight our main contributions. The service structure section presents the design model of the service, which can be monolithic, a set of independent tasks, modular (i.e., a group of tightly-coupled tasks), and microservices (i.e., a set of loosely-coupled tasks). Moreover, it presents the heterogeneity of the constituent parts of IoT services in terms of their computation size and data flow, which we refer to as granular heterogeneity. In the environmental architecture section, tiering is studied, representing the orientation of resources in the environment. It also presents the properties of IoT devices and Edge/Fog servers in terms of their number and heterogeneity. In addition, the environmental architecture studies whether any FSs are able to cooperate for the execution of different IoT services. Also, the multi-cloud column illustrates whether a particular work considers several cloud service providers with heterogeneous resources. The decision engine section depicts the optimization characteristics of each method in terms of the main optimization perspective (i.e., IoT, system, hybrid), problem modeling, and main QoS constraints applied to the problem. Furthermore, this section presents the decision planning technique employed for the offloading/placement, and whether these proposals offer high scalability and adaptability when the computing environment changes.
\par
A fog computing environment is highly stochastic and dynamic, in which many factors affect the search for the optimal offloading solution. Considering the literature, the most complex and heterogeneous fog computing environment comprises multiple heterogeneous IoT devices with diverse services with different QoS requirements, heterogeneous FSs, and heterogeneous multi CSs. Besides, services with dependent constituent parts induce further constraints on the offloading problem. Therefore, traditional approximation techniques, heuristics, and metaheuristics used in several works in the literature, such as \cite{bahreini2021vecman, chen2019collaborative,kimovski2021mobility}, cannot be used efficiently to make offloading decisions in fog computing environments \cite{qi2019knowledge,goudarzi2021distributedDDRL}. Moreover, the high cost of exploration and the low convergence rate of centralized DRL agents in the related literature such as \cite{wang2020fast,zhang2019task,huang2019deep,chen2018optimized,li2022deep,bansal2022urbanenqosplace} form a barrier to the practical deployment of centralized DRL-based decision engines in highly distributed computing environments, especially when the number of features, the complexity of environments, and applied constraints to the offloading problem increase. Finally, there are some DDRL works in the literature, such as \cite{qi2019knowledge,garaali2022learning,zhou2023cost}, that use parameter-sharing DDRL techniques. Although these parameter-sharing techniques are distributed, they do not efficiently use the experience trajectories obtained by different actors because each actor trains its local policies based on its limited experience trajectories and then forwards these parameters to the learner for aggregation/training. Moreover, sharing parameters between actors and learners in parameter-sharing DDRL techniques is more costly, since the size of parameters to be shared among actors and learners is larger than the weights of experience trajectories. To address these issues, we propose a distributed deep reinforcement learning technique, called $\mu$-DDRL, which works based on sharing experience trajectories among actors and learners for offloading services, that consist of heterogeneous tasks with different QoS requirements. We use the asynchronous PPO technique, which employs an actor-learner framework with multiple distributed actors. The actors work in parallel and generate batches of experience trajectories from their own environments that can be shared with the learner. Therefore, the exploration cost of $\mu$-DDRL is significantly reduced, and the batch used in each training iteration is further diversified, leading to a more efficient offloading technique. Also, $\mu$-DDRL uses two important techniques, namely V-trace \cite{espeholt2018impala} and PPO Clipping \cite{schulman2017proximal}, to correct the discrepancy between the learner policy and the actor policies, caused by decoupled acting and learning.
\section{System Model and Problem Formulation}
\label{system}
\begin{figure}[t]
	\centering 
	\includegraphics[width=\linewidth, height=4cm]{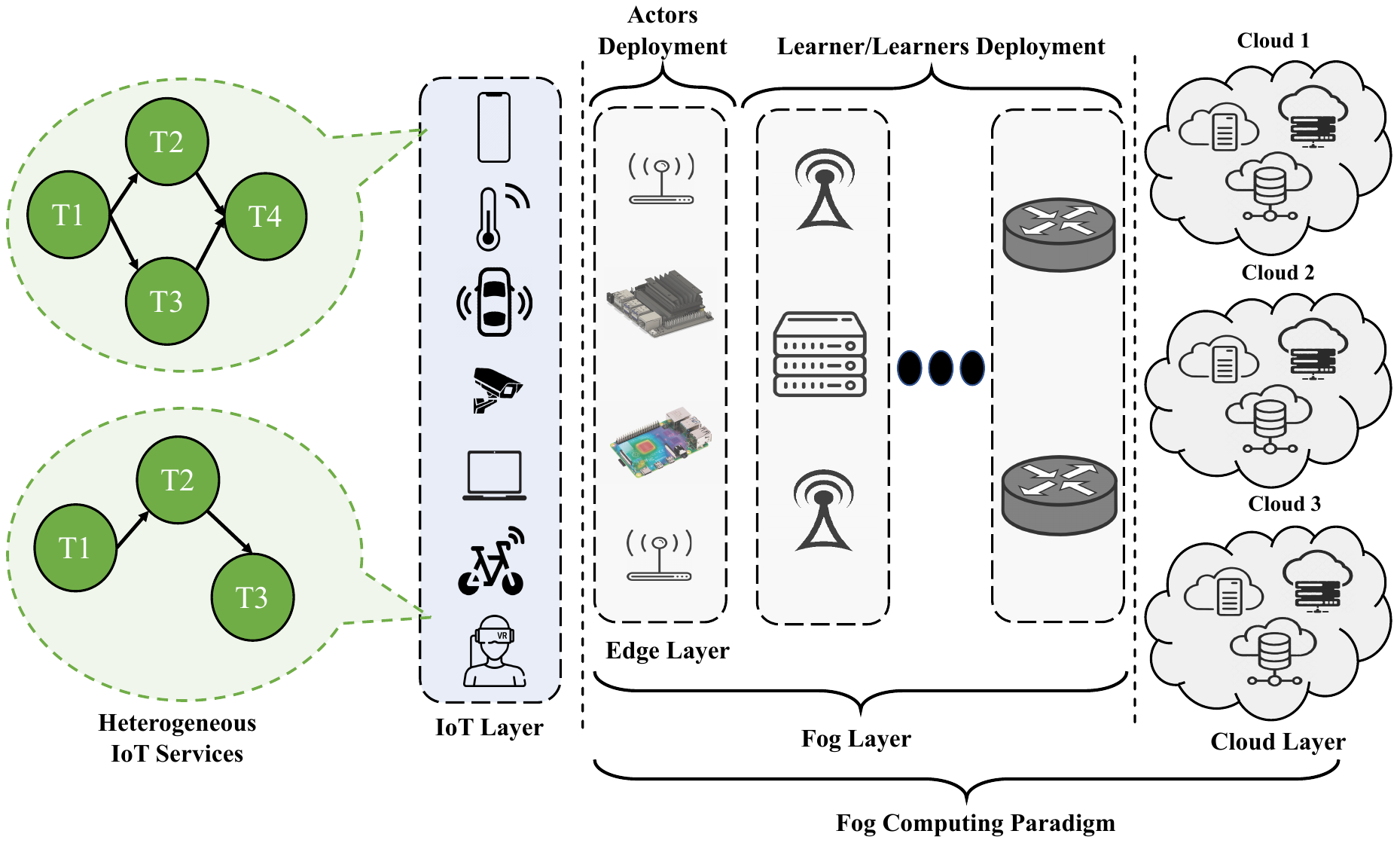}
	\caption{An overview of our system model}
	\label{fig:systemmodel}
\end{figure}

In our system model, we assume that each service may have a different number of tasks with various dependency models, which can be represented as directed cyclic graphs (DAGs) \cite{qi2019knowledge}. Moreover, some predefined servers (i.e., brokers) in proximity of IoT devices are responsible for making service placement decisions based on the resource requirements of each service, its QoS requirements, and the system's properties. These servers are accessible with low latency and high bandwidth, which helps reduce the start-up time of the services. An overview of our system model in fog computing is shown in Fig.~\ref{fig:systemmodel}\footnote{For a detailed description of protocols, technologies, and technical aspects, please refer to https://github.com/Cloudslab/FogBus2}. Appendix~\ref{appendix:paramtersandnotation} presents a summary of the parameters and definitions used.
%
\subsection{Service Model}
We define each service as a DAG $G=(\mathcal{V},\mathcal{E})$, in which $\mathcal{V}=\{v_{i}|1\leq i \leq |\mathcal{V}|\}, |\mathcal{V}|=L$ shows the set of tasks within each service, where the $i$th task is shown as $v_i$. The edges of the graph, $\mathcal{E}=\{e_{i,j}|v_{i}, v_{j} \in \mathcal{V},\thinspace i \neq j\}$, denote the data dependencies between tasks, with $e_{i,j}$ representing a data flow between $v_i$ (parent) and $v_j$ (child). Additionally, the weight of an edge, $e_{i,j}^{w}$, represents the amount of input data that task $v_{j}$ receives from task $v_i$.
\par
A task $v_j$ is represented as a tuple $<v_{j}^{w}$,$v_{j}^{ram}$,$\zeta_{v_j}>$, where $v_{j}^{w}$ is the number of CPU cycles required for the processing of the task, $v_{j}^{ram}$ is the required amount of RAM required to run the task, and $\zeta_{v_j}$ is the maximum tolerable delay for the task (which restricts the execution time of each task). Considering the dependency model between tasks, $\mathcal{P}(v_{j})$ is defined as the set of all predecessor tasks of $v_j$. For each DAG $G$, tasks without parent tasks are called input tasks, while tasks without children are called exit tasks.		

%
\subsection{Problem Formulation}
\label{sec:problem_formula}
The set of available servers is defined as $\mathcal{M}$, where $|\mathcal{M}|=M$. We represent each server as $m^{y,z} \in \mathcal{M}$, in which $y$ illustrates the server's type (i.e., IoT device, FSs, CSs) and $z$ presents the index of the corresponding server's type. Therefore, we define the offloading configuration of task $v_j$ as follows:
\begin{equation}
x_{v_j}=m^{y,z}
\end{equation}
Accordingly, for each service with $L$ tasks, the offloading configuration $\mathcal{X}$ can be defined as:
\begin{equation}
	\mathcal{X}=\{x_{v_j}|v_j \in \mathcal{V},1\leq j \leq L\}
\end{equation}	
Tasks within a service can be sorted in a sequence by an upward ranking algorithm that guarantees each task can only be executed after the successful execution of its predecessor tasks \cite{goudarzi2021distributedDDRL}. Also, the upward ranking algorithm defines a priority for tasks that can be executed in parallel and, accordingly, sorts them \cite{wang2020fast,goudarzi2021distributed}. For parallel tasks, we define $CP(v_i)$ as a function to indicate whether or not each task is on the critical path of the service based on the upward ranking function and the associated computational cost of running tasks \cite{wang2020fast,qi2019knowledge} (i.e., a set of tasks and the corresponding data flow resulting in the highest execution time). Also, $\mathcal{CP}$ presents the set of tasks on the critical path of the service.
\subsubsection{Optimization model}
For each task $v_j$, the execution time can be defined as the time it takes for the input data to become available for that task $\mathscr{T}_{x_{v_j}}^{input}$ plus the processing time of the task on the corresponding server $\mathscr{T}_{x_{v_j}}^{proc}$:

\begin{equation}
\mathscr{T}_{x_{v_j}}= \mathscr{T}_{x_{v_j}}^{proc} + \mathscr{T}_{x_{v_j}}^{input}
\label{exeTimePerTask}
\end{equation}

\noindent
where $\mathscr{T}_{x_{v_j}}^{proc}$ can be obtained based on the CPU cycles required by the task ($v_{j}^{w}$) and the processing speed of the corresponding assigned server $f^{s}_{x_{v_j}}$:

\begin{equation}
\mathscr{T}_{x_{v_j}}^{proc}= \frac{v_{j}^{w}}{f^{s}_{x_{v_j}}}
\end{equation}

\noindent
 $\mathscr{T}_{x_{v_j}}^{input}$ is estimated as the time it takes for all input data to arrive to the server assigned to $v_j$  (i.e., $x_{v_j}$) from its predecessors:
\begingroup
\footnotesize	
\begin{equation}
\mathscr{T}_{x_{v_j}}^{input}= \max((\frac{e_{i,j}^{w}}{b(x_{v_i},x_{v_j})}+l(x_{v_i},x_{v_j}))\times SS(x_{v_i},x_{v_j})),\; \forall v_i \in \mathcal{P}(v_j)
\end{equation}
\endgroup 

\noindent
in which $b(x_{v_i},x_{v_j})$ illustrates the data rate (i.e., bandwidth) between the selected servers for the execution of $v_i$ and $v_j$, respectively. Moreover, $l(x_{v_i},x_{v_j})$ depicts the communication latency between two servers (i.e., an IoT device and a remote server or two remote servers), and is calculated based on the propagation speed for the communication medium (i.e., $\eth^c$) and the Euclidean distance between the coordinates of the participating servers (i.e., $d(x_{v_i},x_{v_j})$) in the Cartesian coordinate system:
\begin{equation}
\label{eq:latency}
l(x_{v_i},x_{v_j})=\frac{d(x_{v_i},x_{v_j})}{\eth^c}	
\end{equation}
where $d(x_{v_i},x_{v_j})$ is calculated as follows:
\begin{equation}
d(x_{v_i},x_{v_j})=\sqrt{(x_{v_i}^x-x_{v_j}^x)^2+(x_{v_i}^y-x_{v_j}^y)^2}
\end{equation}
The $SS(x_{v_i},x_{v_j})$ is a function that indicates whether the assigned servers to each pair of tasks are the same (i.e., $0$ if $x_{v_i} = x_{v_j}$) or different (i.e., $1$ if $x_{v_i} \neq x_{v_j}$). Because the fog computing environment is dynamic, heterogeneous, and stochastic, $f^{s}_{x_{v_j}}$, $b(x_{v_i},x_{v_j})$, and $l(x_{v_i},x_{v_j})$ may have different values from time to time.
\par		
The principal optimization objective is minimizing the execution time of each service by finding the best possible configuration of surrogate servers for the execution of the service's tasks, as defined below.

\begin{equation}
\label{globalOptimization}
\min (\mathcal{T}(\mathcal{X}))
\end{equation}

where
\begin{equation}
\mathcal{T}(\mathcal{X}) = \sum\limits_{j=1}^{L}CP(v_{j})\times \mathscr{T}_{x_{v_j}}
\label{timeModel}
\end{equation}

$s.t.$
\vspace{-0.3cm}
\begin{eqnarray}
&&CS1:\;S_n(x_{v_j})=1,\; \forall x_{v_j} \in \mathcal{X}\\
&&CS2:\;\mathcal{T}(v_i) \leq \mathcal{T}(v_i+ v_j),\; \forall v_{i}\in\mathcal{P}(v_{j})\\
&&CS3:\; \sum\limits_{\forall v_j \in \mathcal{V}} IA(v_j,m^{y,z}) \times v_j^{ram}  \leq R(m^{y,z})\\
&&CS4:\; \mathscr{T}_{x_{v_j}} \leq \zeta_{v_j}, \; \forall v_j \in \mathcal{V}
\end{eqnarray}
\noindent
\noindent
where $CP(v_j)$ can be obtained from:
\begin{eqnarray}
\label{eq.criticalpathcondition}
CP(v_j)= \left\{ \begin{tabular}{cc} $1$, & $v_j \in \mathcal{CP}$ \vspace{0.1cm}\\	
 			$0$, & $otherwise$
\end{tabular}\right.
\end{eqnarray}

$CP(v_j)$ is equal to $1$ if the task $v_j$ is on the critical path and $0$ otherwise. $CS1$ restricts the assignment of each task to exactly one server at a time. Also, $CS2$ shows that the cumulative execution time of $v_j$ is always larger or equal to the execution time of its predecessors' tasks \cite{xu2019computation}, which guarantees the execution order of tasks with data dependency. $CS3$ states that the summation of the required RAM for all tasks assigned to one server should always be less than or equal to the available memory on that server $R(m^{y,z})$. $IA(v_j,m^{y,z})$ is an indicator function to check if the task $v_j$ is assigned to the server $m^{y,z}$ ($IA=1$) or not ($IA=0$). Finally, $CS4$ states that the execution time of each task should be within the range of the task's tolerable delay.
\par
The service offloading problem in heterogeneous computing environments is an NP-hard problem \cite{qiu2020distributed,goudarzi2021distributedDDRL}, and the complexity of the problem grows exponentially as the number of servers and/or tasks within a service increases. Therefore, the optimal solution to the service offloading problem cannot be obtained in polynomial time. 
\section{DRL Model}
\label{sec:DRLModel}
In the DRL, reinforcement learning and deep learning are combined to transform a set of high-dimensional inputs into a set of outputs to make decisions. In DRL, the learning problem is formally modeled as a Markov Decision Process (MDP) which is mainly used to make decisions in stochastic environments. We define a learning problem by a tuple $<\mathbb{S},\mathbb{A},\mathbb{P},\mathbb{R},\gamma>$, where $\mathbb{S}$ shows the state space and $\mathbb{A}$ presents the action space. The probability of a state transition between states is shown by $\mathbb{P}$. Finally, $\mathbb{R}$ and $\gamma \in [0,1]$ denote the reward function and discount factor, respectively. Also, the continuous time horizon is divided into multiple time steps $t \in \mathbb{T}$. In each time step $t$, the DRL agent interacts with the environment and receives the state of the environment $s_t$. Consequently, the agent uses its policy $\pi(a_t|s_t)$ and chooses an action $a_t$. The agent performs the action $a_t$ and receives a reward in that time step $r_t$. The agent aims at maximizing the expected total of future discounted rewards:
\begin{equation}
\mathbb{V}^{\pi}(s_t) = \mathbb{E}_{\pi} [\sum\limits_{t \in \mathbb{T}}\gamma^{t}r_t]
\end{equation}
\noindent
in which $r_t = \mathbb{R}(s_t, a_t)$ represents the reward at time step $t$, and $a_t \thicksim \pi(.|s_t)$ is the action at $t$ when following the policy $\pi$. Furthermore, because DNN is used to approximate the function, $\theta$ shows the corresponding parameters.
\par
In what follows, the main concepts of the DRL for the service offloading problem in the fog computing environment are described:
\begin{itemize}
\item \textbf{State space $\mathbb{S}$:} The state is defined as the agent's observation of fog computing environments. Accordingly, we represent the state space as a set of characteristics for all servers, IoT devices, and corresponding services in different time steps: 
\begin{equation}
\mathbb{S}=\{s_t|s_t = (F^{\mathcal{M}}_{t}, F^{v_j}_{t}), \forall t \in \mathbb{T}\}
\label{System_State}
\end{equation}
\noindent where $s_t$ shows the system state at time step $t$, $F^{\mathcal{M}}_{t}$ represents the feature vector of all $M$ servers at time step $t$, and $F^{v_j}_{t}$ presents the feature vectors of the current task in a service. The $F^{\mathcal{M}}_{t}$ includes the servers and devices' properties, such as their positions, number of CPU cores, corresponding CPU frequency speed, data rate, ram, etc. If each server/device has a maximum of $K_1$ features, we can define the feature vector of all $M$ servers/devices at time step $t$ as:
\begin{equation}
F^{\mathcal{M}}_{t} = \{f_i^{m^{y,z}}|\forall m^{y,z} \in 	\mathcal{M},1\leq\ i \leq k_1\} 
\label{FV_servers}
\end{equation}
\noindent
in which $f_i^{m^{y,z}}$ represents the $i$th feature corresponding to the server $m^{y,z}$. Also, the $F^{v_j}_{t}$ includes the features of a current task $,v_j$, such as required computation, RAM, dependency model of the corresponding tasks, and service offloading configuration of prior tasks, just to mention a few. As tasks within a service are sorted by upward ranking, the current tasks' dependencies are already solved. Supposing that each task at maximum contains $k_2$ features, we can define the feature vector of task $v_j$ at time step $t$ as:
\begin{equation}
F^{v_j}_{t} = \{f_i^{v_j}| v_j \in \mathcal{V}, \forall i \; 1\leq i \leq k_2\} 
\label{FV_tasks}
\end{equation}		    
\noindent
in which $f_i^{v_j}$ shows the $i$th feature of the task $v_j$.
\item \textbf{Action space $\mathbb{A}$:} Since we aim to find the best configuration of servers for tasks within a service, the action space $\mathbb{A}$ is related to all available servers, defined in what follows:
\begin{equation}
\mathbb{A} = \mathcal{M}
\end{equation}
\noindent Also, each action in time step $t$, i.e., $a_t$, is defined in the assignment of a server to the current task:
\begin{equation}
a_t = x_{v_j}= m^{y,z}
\end{equation}
\noindent

%
\item \textbf{Reward function $\mathbb{R}$:} As presented in Eq.~\ref{globalOptimization}, the optimization goal is to minimize the execution time of each service while satisfying the QoS. Hence, the reward function per time step $t$ is defined as the negative value of Eq.~\ref{exeTimePerTask} if the task can be properly executed within the maximum tolerable delay (i.e., $\mathscr{T}_{x_{v_j}} \leq \zeta_{v_j}$). As tasks are currently sorted based on the upward-ranking function, tasks that incur the highest execution cost (i.e., tasks on the critical path of the service) receive higher priority. Thus, tasks incurring the highest execution cost can be assigned to more powerful servers at the initial stages of the offloading decision-making process. Although it is not obligatory for the step-reward function to directly align with the long-term reward for the DRL agent, enhancing their correlation can enhance the agent's overall performance. In our reward function, the step reward is correlated to the long-term reward because it tries to reduce the execution time of each task, ultimately resulting in the reduction of the overall service execution time. Moreover, $\Phi$ is defined as a failure penalty to penalize actions that violate the maximum tolerable delay for that task for any reason. Accordingly, $r_t$ is defined as:

\begin{eqnarray}
\label{eq.rewardFunctionTimeStep}
r_t= \left\{ \begin{tabular}{cc} $-\mathscr{T}_{x_{v_j}}$, & $\mathscr{T}_{x_{v_j}} \leq \zeta_{v_j}$ \vspace{0.1cm}\\	
 			$\Phi$, & $otherwise$
\end{tabular}\right.
\end{eqnarray}

\end{itemize}			
\section{$\mu$-DDRL: Distributed DRL Framework}
\label{sec:placement}
In this section, we mainly describe the $\mu$-DDRL framework, which is a technique for high-throughput distributed deep reinforcement learning based on the actor-critic APPO architecture. $\mu$-DDRL is designed for QoS-aware service offloading and aims to minimize the execution time of distinct services in highly heterogeneous and stochastic computing environments.
\par
\begin{figure*}[t]
\centering 
\includegraphics[width=0.8\linewidth, height=5cm, trim=0.1in 0.25in 0in 0.25in]{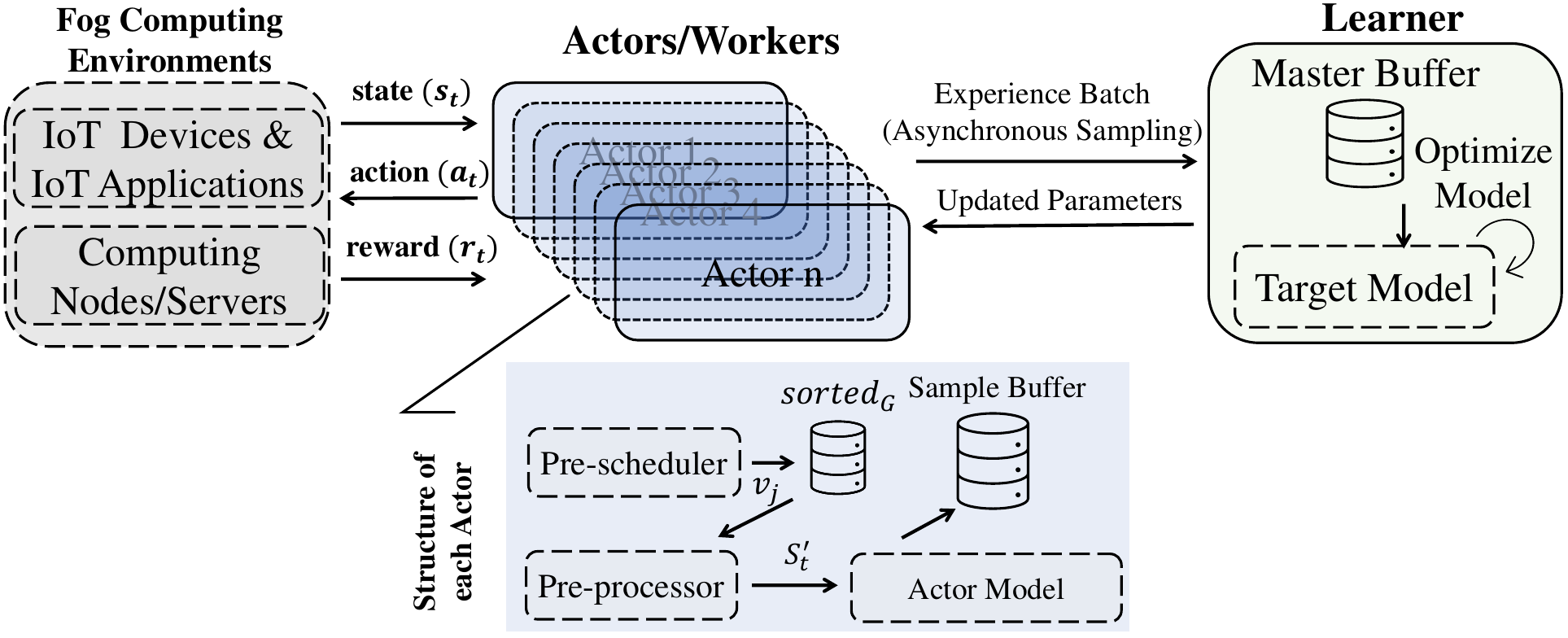}
\caption{An overview of $\mu$-DDRL framework}
\label{fig:mu-DDRL}
\end{figure*}  
In the actor-critic architecture, the policy, $\pi(a_t|s_t;\theta)$ is parameterized by $\theta$. Gradient ascent on the variance of the expected total future discounted reward (i.e., $\sum\limits_{k=0}^{\infty}\gamma^{k}r_{t+k}$) and the learned state-value function under policy $\pi$ (i.e., $\mathbb{V}^{\pi}(s_t)$) are used to update $\theta$. The actor interacts with its environment and obtains the state $s_t$ and selects an action $a_t$ according to the parameterized policy $\pi(a_t|s_t;\theta)$, resulting in a reward $r_t$ and the next state $s_{t+1}$. The critic utilizes the obtained rewards to evaluate the current policy based on the Temporal Difference (TD) error between the current reward and the estimation of the value function. Different DNNs are used as function approximators for the actor and the critic, which are trained independently. The parameters of the actor network are updated by feedback from the TD error to improve the selection probability of suitable actions. Additionally, the parameters of the critic network are updated to obtain a higher quality value estimation. The decoupled act and learning in actor-critic architecture makes this method a suitable option for long-term performance optimizations in distributed computing environments. 
\paragraph*{\textbf{APPO and its role in $\mu$-DDRL}} While the actor-critic architecture is suitable for highly distributed environments, it still suffers from slow learning speed and high exploration costs. APPO is a distributed DRL architecture that helps significantly parallelize the collection of the distributed experience trajectories from different actors and achieves very high throughput. The distributed experience-sharing approach, in which actors generate heterogeneous trajectories in parallel and share their experience trajectories with learners/learners, can significantly improve the learning speed and exploration costs of actor-critic frameworks (and DRL frameworks in general). $\mu$-DDRL works based on the decoupled actor-critic and APPO. Hence, each actor in the distributed fog computing environments makes service offloading decisions and creates experience trajectories. Each actor asynchronously forwards its collection of experience trajectories to the learner for training. Hence, the experience trajectories of slow actors (actors generating fewer experience trajectories) and fast actors (actors generating more experience trajectories) can be mixed together in the learner, resulting in more diversified experience trajectories. Moreover, when the learner updates the policy, the local policy of slow and fast actors will be updated together, so that each actor can more efficiently make service offloading decisions. Fig.~\ref{fig:mu-DDRL} presents an overview of the $\mu$-DDRL framework. In what follows, the roles of each actor and critic (i.e., learner) are described in detail.
\subsection{$\mu$-DDRL: Actor Role}
In the $\mu$-DDRL framework, actors act as brokers in the environment that make offloading decisions using their own local policy $\kappa$ for incoming service offloading requests from smart devices (e.g., smart transportation cameras). In a fog computing environment, where computational resources are distributed, actors can be deployed on distributed and heterogeneous servers in the environment. Moreover, distributed actors can work in parallel and interact with the environment to make service-offloading decisions. This distributed behavior reduces the incoming requests' queuing time, which improves the service-ready time for users. Algorithm~\ref{alg:actor} presents how each actor/broker interacts with the environment, makes service offloading decisions, and builds experience trajectories.
\par
Initially, the local policy of each actor $\kappa$ is updated by the policy of the learner $\pi$, to ensure that each actor works with the most optimized and updated policy to make service offloading decisions (line 3). Then, each actor uses the local policy $\kappa$ to make $N$ offloading decisions. The actor fetches the metadata of a service request from $S_Q$ whenever it begins making the offloading decision for a new service $G$ (line 7). Next, the metadata of the new service $G$ is fed into the \textit{Pre-sch()} function, which sorts the tasks within each service while satisfying the dependency constraints among tasks considering the DAG of each service and outputs the list of sorted tasks, $sort_G$ (line 8). The \textit{Pre-sch()} function works based on the upward ranking to sort tasks within a service \cite{goudarzi2021distributedDDRL,wang2020fast}. While other sorting algorithms, such as topological sort, can also satisfy the dependency constraints of tasks within a DAG, these sorting algorithms, by default, do not provide any order for tasks that can be executed in parallel. The upward ranking algorithm not only defines an order for the parallel tasks but also helps identify the critical path of a DAG-based service by estimating the average execution cost of each task and its related edges on different servers. Next, the DAG of service $G$, list of available servers $\mathcal{M}$, and sorted lists of tasks $sort_G$ are given as input to the \textit{RCVInitState()} function to build the system state $s_i$ (line 9). Afterwards, the state of $flag_{in}$ is changed to $False$ until the actor starts making a decision for a new incoming service requiring the re-execution of \textit{Pre-sch()} to obtain a new sequence of ordered tasks (line 10). Otherwise, the current state $s_i$ of the system is obtained by \textit{RCVCurrState()} function based on Eq.~\ref{System_State}, which contains the feature vectors of the servers $F^{\mathcal{M}}_{t}$ and the current task of service $F^{v_j}_{t}$ (line 12). The \textit{Pre-proc()} function receives the feature vector of the current state $s_i$ and normalizes the values (line 14). Next, the actor performs the service offloading based on the current local policy $\kappa$ and outputs the action $a_i$ using the \textit{SOEngine()} function (line 15). The action $a_i$ is then executed, which means that the current task is offloaded onto the assigned surrogate server. The reward of this offloading action is then calculated by the \textit{StepCostCal()} function based on Eq.~\ref{eq.rewardFunctionTimeStep}, which takes into account the execution of the task on the assigned server within the specified maximum tolerable delay $\zeta_{v_j}$ (line 16). The \textit{BuildNextState()} function then prepares the next state of the environment, and the experience batch $EB$ of the current actor is updated by the experience tuple ($s_i$,$a_i$,$r_i$,$s_{i+1}$) (lines 17-18). If the service offloading for the current service is finished, the actor calculates the total execution cost of the current service using \textit{TotalCostCal()} function based on Eq.~\ref{globalOptimization} and adjusts the state of $flag_{in}$ (lines 19-20). This process is continuously repeated for the $N$ steps, after which the actor sends the experience batch to the learner and restarts the process with an updated local policy $\kappa$ (line 24).

\begin{algorithm}[!t]
\footnotesize
\caption{Each actor's role} \label{alg:actor}
\SetKwData{Left}{left}
\SetKwData{This}{this}
\SetKwData{Up}{up}
\SetKwFunction{Union}{Union}
\SetKwFunction{FindCompress}{FindCompress}
\SetKwInOut{Input}{Input}
\SetKwInOut{Output}{Output}
\SetKwInOut{Parameter}{Parameter}
\Input{$\pi$: The policy of the learner}
\tcc{$N$:Maximum steps's number, $\kappa$: the local policy of each actor, $EB$: experience batch, $S_{Q}$: Queue of received requests, $G$: current service}
$flag_{in}$=True\\
\While {$True$}{
    $\kappa$=UpdateLocalPolicy($\kappa$, $\pi$)\\
    $i=0$ \\
    \While{$i<N$}{
        \eIf{$flag_{in}$=True}{
            $G$=$S_{Q}$.dequeue()\\
            $sort_G$ = Pre-sch ($G$) \\
            $s_i$=RCVInitState($G$, $\mathcal{M}$, $sort_G$) \\
            $flag_{in}$=False\\
        }
        {$s_i$=RCVCurrState() \\}
        $s_i$=Pre-proc($s_i$) \\
        $a_i$=SOEngine($s_i$, $\kappa$) \% Service Offloading Engine\\
        $r_i$=StepCostCal($s_i$, $a_i$)  \% $\rightarrow$ Eq.~\ref{eq.rewardFunctionTimeStep} \\
        $s_{i+1}$ = BuildNextState($s_{i}$, $a_i$)\\
        $EB$.update($s_i$, $a_i$, $r_i$, $s_{i+1}$)\\
        \If{Finish($G$)}{
            TotalCostCal($G$) \% $\rightarrow$ Eq.~\ref{globalOptimization}\\ 
            $flag_{in}$=True\\
        }			
    }
    ShareExpeienceToLearner($EB$)\\ 
    
}

\end{algorithm}
\subsection{$\mu$-DDRL: Learner Role}
In the $\mu$-DDRL framework, the main roles of the learner are managing the incoming experience batches from different actors, training and updating the target policy $\pi$, and updating the actor policies $\kappa$.
\paragraph*{\textbf{V-trace, PPO Clipping, and their role in $\mu$-DDRL}} In asynchronous RL techniques, a policy gap between the actor policy $\kappa$ and learner policy $\pi$ may happen. Generally, two major types of techniques are designed to cope with off-policy learning \cite{petrenko2020sample}. First, applying trust region methods by which the technique can stay closer to the behavior policy (i.e., the policy used by actors when generating experiences) during the learning process, which results in higher quality gradient estimates obtained using the samples from this policy \cite{schulman2017proximal}. The second type is employing importance sampling to correct targets for the value function to improve the approximation of the discounted sum of the rewards under the learner policy. In this work, we use PPO Clipping \cite{schulman2017proximal} as a trust region technique and the V-trace algorithm \cite{espeholt2018impala} as an importance sampling technique, which uses truncated importance sampling weights to correct the value targets. It has been shown that these techniques can be applied independently since V-trace corrects the training objective and PPO Clipping protects against destructive parameter updates. Overall, the combination of PPO-Clipping and V-trace leads to more stable training in asynchronous RL \cite{petrenko2020sample}.
\par
Algorithm~\ref{alg:learner} presents the learner role in $\mu$-DDRL. The learner actively interacts with the actors and asynchronously receives the experience trajectories $EB_a$. The learner has a master buffer, called $MB$, which stores the incoming trajectories from actors (line 5). Whenever the size of the master buffer $MB$ reaches the size of the training batch $TB$, the \textit{BuildTrainBatch()} function fetches sufficient samples from the master buffer $MB$ for training (lines 6-10). Next, the \textit{OptimizeModel()} function computes the policy, computes value network gradients, and updates the policy and value network weights, as follows \cite{luo2019impact}:


\begin{eqnarray}
\label{eq:Jthetha}
\nabla_{\theta} J(\theta)=\frac{1}{|TB|} \sum_{j \in TB} \min (\mathbb{Z}\times \hat{A}_{\mathbb{V}(\operatorname{GAE})},Clip(Z,\epsilon)\hat{A}_{\mathbb{V}(\operatorname{GAE})})
\end{eqnarray}

\noindent
where $\mathbb{Z}$ is $\frac{\pi_t}{\kappa}$, $Clip(.,\epsilon)=Clip(.,1-\epsilon,1+\epsilon)$ is a clipping function, $TB$ shows the training batch, $\pi_{t}$ shows the target policy, and the V-trace GAE-$\lambda$ (i.e., $\hat{A}_{\mathbb{V}(\operatorname{GAE})}$) modifies the advantage function by adding clipped importance sampling terms to the summation of TD errors:

\begin{equation}
\label{eq:vtraceLambda}
\hat{A}_{\mathbb{V}(\operatorname{GAE})}=\sum_{i=t}^{t+n-1}(\lambda \gamma)^{i-t}\left(\prod_{j=t}^{i-1} c_j\right) \delta_i \mathbb{V}
\end{equation}

\noindent
where $\delta_i \mathbb{V}$ is the importance sampled TD error introduced in V-trace \cite{espeholt2018impala}, defined as follows:

\begin{equation} 
\label{eq:vtraceTD}
\delta_i\mathbb{V} = \rho_i(r_i+\gamma \mathbb{V}(s_{i+1})-\mathbb{V}(s_i))
\end{equation}    
\noindent

\noindent
The $\gamma$ is a discount factor. Moreover, $c_j=min(\overline{c},\frac{\pi_{t}(a_j|s_j)}{\kappa(a_j|s_j)})$ and $\rho_i=min(\overline{\rho},\frac{\pi(a_i|s_i)}{\kappa(a_i|s_i)})$ are clipped Importance Sampling (IS) weight in V-trace, controlling the speed of convergence and the value function to which the technique converges, respectively. Finally, the value network gradients are computed as follows.

\begin{equation}
\label{eq:networkGradients}
\nabla_w L(w)=\frac{1}{|TB|} \sum_j\left(\mathbb{V}_w\left(s_j\right)-\hat{\mathbb{V}}_{\mathbb{V}(\operatorname{GAE})}\left(s_j\right)\right) \nabla_w \mathbb{V}_w\left(s_j\right)
\end{equation}

\noindent
Next, the policy and value network weights are updated. Finally, the learner sends the updated policy and weights to the actors (line 12). Taking into account the target network stored in the learner, the \textit{OptimizeModel()} function has the ability to run multiple stochastic gradient descent/Ascent (SGD/SGA). Moreover, $\mu$-DDRL offers a highly scalable solution because as the number of actors increases, the number of shared experiences with the learner increases, and also the training batch is further diversified. As a result, distributed DRL agents can learn faster and adapt themselves to stochastic fog computing environments. In addition, the asynchronous nature of acting and learning in $\mu$-DDRL helps newly joined actors in the environment to initialize their policy and parameters with the most recent and optimized policy of the learner rather than start making service offloading decisions using the naive actor policy, resulting in better offloading decisions.

\begin{algorithm}[!t]
\footnotesize
\caption{Learner's role} \label{alg:learner}
\SetKwData{Left}{left}
\SetKwData{This}{this}
\SetKwData{Up}{up}
\SetKwFunction{Union}{Union}
\SetKwFunction{FindCompress}{FindCompress}
\SetKwInOut{Input}{Input}
\SetKwInOut{Output}{Output}
\SetKwInOut{Parameter}{Parameter}
\Input{$EB_{a}$: Actors' batch of experience}
\tcc{$list_{a}$: actors' list, $\pi$: the policy of the learner, $MB$: master buffer, $TB$: training batch}
\While {True}{
    $flag_{tr}$=False\\
    $MB$=$\emptyset$\\
    \While{$flag_{tr}$==False}{
        $MB$.add($EB_{a}$)\\
        \If{$size(TB)\leq size(MB)$}{
            TB=BuildTrainBatch($MB$)\\ 
            $flag_{tr}$==True
        }
    }
    OptimizeModel($TB$) \% $\rightarrow$ Eqs.~\ref{eq:Jthetha}, \ref{eq:vtraceLambda}, \ref{eq:vtraceTD}, \ref{eq:networkGradients}
    \\
    UpdateActors($list_{brokers}$)			
}
\end{algorithm}
\section{Performance Evaluation}
\label{sec:evaluation}
In this section, we describe the evaluation setup and baseline techniques used for the performance evaluation. Then, the hyperparameters of $\mu$-DDRL are illustrated and the performance of $\mu$-DDRL and its counterparts are evaluated.
\subsection{Evaluation Setup}
OpenAI gym \cite{brockman2016openai} is used to develop a simulation environment for a heterogeneous fog computing environment. The environment comprises Smart Cameras in Transportation Systems (SCTS) as IoT devices with heterogeneous service requests, heterogeneous resource-limited FSs, and powerful CSs. For SCTS' resources, we consider a 1 GHz single-core CPU. As heterogeneous FSs, we consider 30 FSs, where each FS has a CPU with four cores with 1.5-2GHz processing speed and 1-4GB of RAM\footnote{The resources of FSs are aligned with resources of Raspberrypi and Nvidia Jetson platform}. Moreover, we consider 20 CSs, where each CS has an eight-core CPU with 2-3Ghz processing speed and 16-24GB of RAM for CSs. The bandwidth (i.e., data rate) among different servers and SCTS devices is set based on the values profiled from the testbed setup \cite{deng2021fogbus2}. Accordingly, the bandwidth between SCTS devices and FSs is randomly selected between 10-12MB/s, while the bandwidth between SCTS devices and FSs to CSs is randomly selected between 4-8MB/s, similar to \cite{wu2019efficient,goudarzi2021distributedDDRL}. The latency between servers and SCTS devices depends on the respective coordinates of the SCTS devices and remote servers, as defined in Eq.~\ref{eq:latency}.
\par
\begin{figure}[t]
	\centering 
	\includegraphics[width=\linewidth, height=4cm]{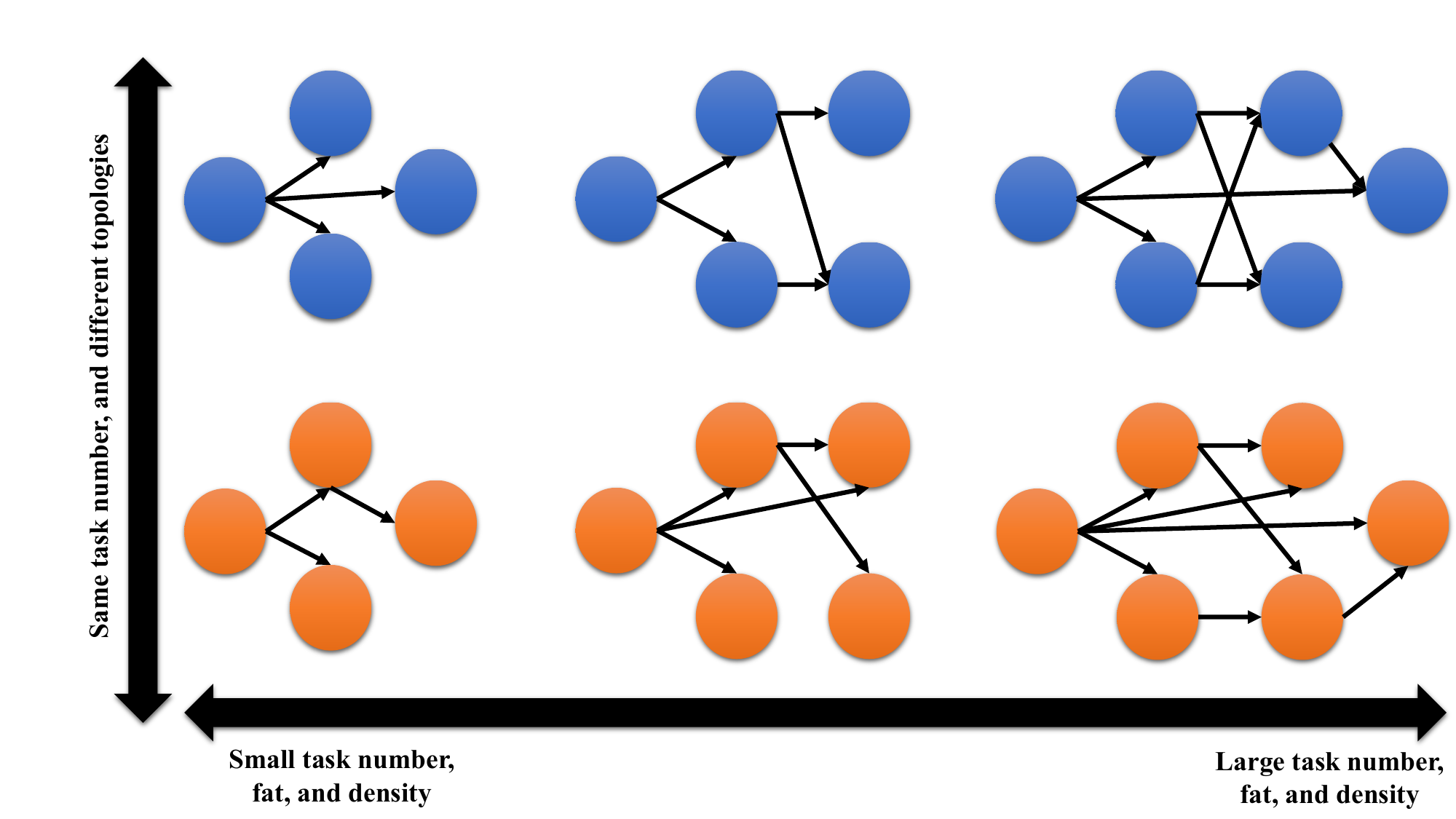}
	\caption{Role of density, fat, and number of tasks in dataset}
	\label{fig:datasetServicefatanddensity}
\end{figure}

Several complex services based on DAG for SCTS devices, which contain heterogeneous tasks with different dependency models, are designed to represent heterogeneous services with different preferences, similar to \cite{qi2019knowledge,wang2020fast}. The service property data contains the number of tasks, the required CPU cycles ($v_{j}^{w}$), RAM usage per task ($v_{j}^{ram}$), the dependency model between tasks, the amount of data transmission between a pair of tasks ($e_{i,j}^{w}$) and the maximum tolerable delay for each task ($\zeta_{v_j}$). Thus, in order to generate heterogeneous DAGs, two important points must be considered. First, the topology of the DAGs depends on the number of tasks within the service ($L$), fat (identifies the width and height of the DAG), and density (controls the number of edges within the DAG). The second point is the weights assigned to each task and edge within each DAG. To generate different DAG topologies, we assume that the number of tasks for different services ranges from 5 to 50 tasks (i.e., $L \in \{5,10,15,20,25,30,35,40,45,50\}$) to show a heterogeneous number of tasks for services \cite{qi2019knowledge,wang2020fast,goudarzi2021distributedDDRL}. Moreover, $fat \in \{0.4,0.5,0.6,0.7,0.8\}$, and $density \in \{0.4,0.5,0.6,0.7,0.8\}$ to create different dependency models between tasks of each service. Taking into account these values, we have generated 250 different topologies for DAGs, so that each DAG presents the potential topology of a service. Fig.~\ref{fig:datasetServicefatanddensity} shows the role of fat, density, and $L$ in the creation of DAGs. Next, for each DAG topology, we generate 100 DAGs with different sets of weights, where $v_{j}^{w}$ ranges from $10^7$ cycles to $3*10^8$ cycles \cite{wang2020fast,dinh2017offloading}, $v_{j}^{ram}$ ranges from 25MB to 100MB \cite{goudarzi2021resource}, $e_{i,j}^{w}$ ranges from 50KB to 2000KB \cite{wang2020fast,cheng2021multi}, and $\zeta_{v_j}$ is randomly selected from $[25-100]$ms \cite{xie2006scheduling,gazori2020saving,cheng2021multi,chen2021locus}. Consequently, 250 different DAG topologies and 100 different weighting configurations per DAG topology result in 25,000 service DAGs with heterogeneous $L$, $fat$, and $density$, and weights. Among these DAGs, 80\% have been used for training $\mu$-DDRL and baseline techniques, while 20\% of the DAGs are used for evaluation. 
\par
The performance of $\mu$-DDRL is compared with the following baseline techniques:
\begin{itemize}
    \item MRLCO: The improved version of the technique proposed in \cite{wang2020fast} is considered for evaluation. This technique is extended to be used in heterogeneous fog computing environments where several IoT devices, FSs, and CSs are available. Also, the reward function is updated to support deadline constraints. This technique uses synchronous PPO as its DRL framework while the networks of the agent are wrapped by the RNN. Furthermore, the agent is tuned based on the hyperparameters used in \cite{wang2020fast}.
    \item Double-DQN: In the literature, several works such as \cite{zhang2019task,chen2018optimized,huang2019deep} have used standard deep Q-learning (DQN) or double-DQN as the agent. Here, we used the optimized Double-DQN technique with an adaptive exploration which provides better efficiency for the agent. The agent model is tuned based on the hyperparameters used in \cite{zhang2019task}.
    \item DRLCO: The enhanced version of the method introduced in \cite{zhou2023cost} is considered for evaluation. Furthermore, we updated the reward function in this work to accommodate heterogeneous services with different topologies and deadline constraints. This approach uses A3C as its underlying DDRL framework.

\end{itemize}
\subsection{Performance Study}
This section describes the $\mu$-DDRL hyperparameters and presents the performance of our proposed technique and its counterparts. 
\subsubsection{$\mu$-DDRL Hyperparameters}
In $\mu$-DDRL, the DNN contains two fully connected layers, and the DNN structure of all agents is exactly the same. For hyperparameter tuning, we performed a grid search and, accordingly, the learning rate $lr$, discount factor $\gamma$, and gradient steps are set to $0.01$, $0.99$, and $2$, respectively. The control parameters of the V-trace, $\overline{\rho}$ and $\overline{c}$, are set to $1$. In addition, the clipping constant $\epsilon$ and the GAE discount factor are set to $0.2$ and $0.95$, respectively. The summary of the hyperparameter configuration is presented in Table~\ref{tab:hyperparameters}.
\begin{table}[!t]
\footnotesize
\caption{Hyperparameters}
\centering
\label{tab:hyperparameters}
\resizebox{1\linewidth}{!}{%
\renewcommand{\arraystretch}{1.2}
\begin{tabular}{|c|c|c|c|}
    
    \hline
    \textbf{Parameter}     & \textbf{Value} & \textbf{Parameter}                               & \textbf{Value} \\ \hline
    FC layers & 2              & Learning Rate $lr$                               & 0.01           \\ \hline
    Gradient Steps            & 2             & Discount Factor $\gamma$                         & 0.99           \\ \hline
    Optimization Technique    & Adam           & V-trace $\overline{\rho}$ & 1              \\ \hline
    Activation Function    & Tanh           & V-trace $\overline{c}$                    & 1              \\ \hline
    Clipping Constant $\epsilon$     & 0.2            &     GAE Discount Factor $\lambda$                  &  0.95\\\hline
\end{tabular}
}
\end{table}
\subsubsection{Execution Time Analysis}	
\begin{figure*}[]
    \begin{subfigure}{.5\textwidth}
        \centering
        \includegraphics[width=\linewidth,height=4cm]{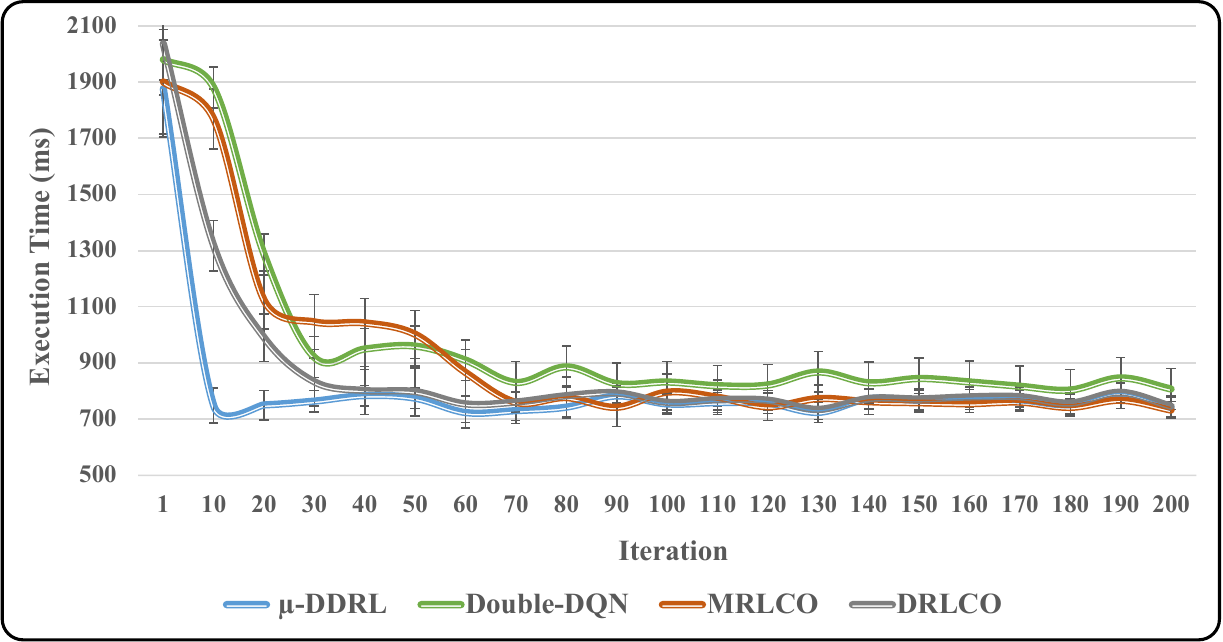}
        \captionsetup{justification=centering}
        \caption{Training dataset: $L \in \{5,10,20,25,30,35,40,45,50\}$, Evaluation dataset: $L=15$}
        \label{fig:ExecutionTimeIteration_15microservices}
    \end{subfigure}%
    \hspace{.02cm}
    \begin{subfigure}{.5\textwidth}
        \centering
        \includegraphics[width=\linewidth,height=4cm]{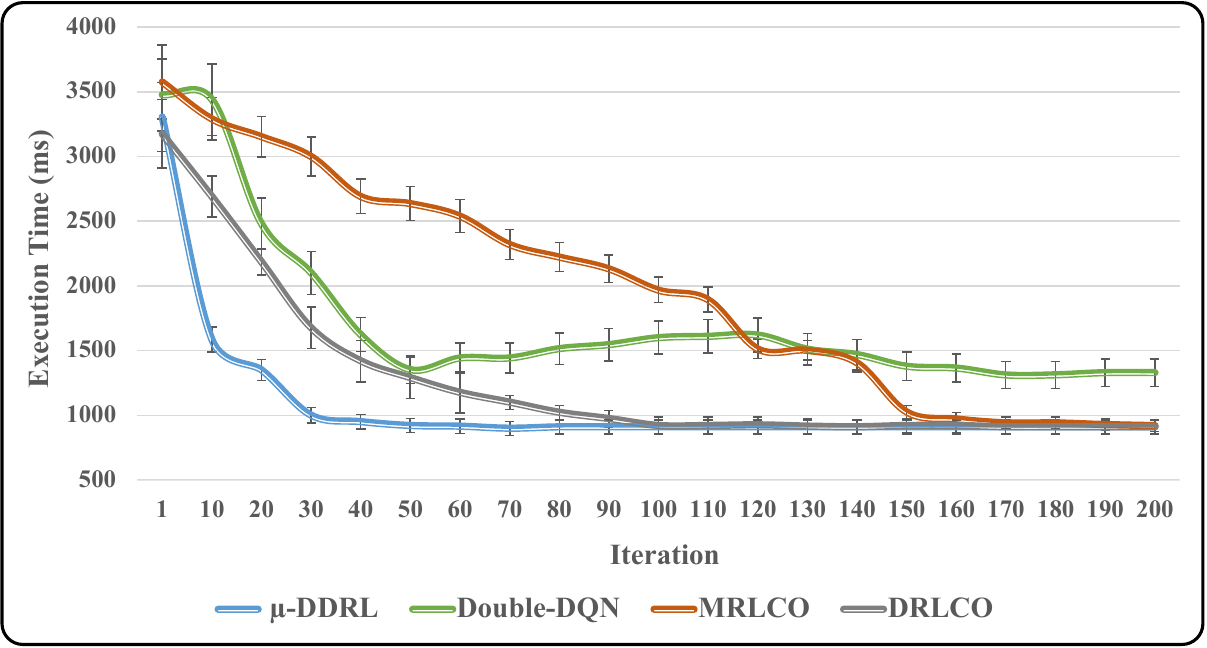}
        \captionsetup{justification=centering}
        \caption{Training dataset: $L \in \{5,10,15,20,25,30,35,45,50\}$, Evaluation dataset: $L=40$}
        \label{fig:ExecutionTimeIteration_40microservices}
    \end{subfigure}
    
    \caption{Execution time analysis vs number of target policy updates}
    \label{fig:ExecutionTimeIteration}
\end{figure*}
This experiment demonstrates the performance of different service offloading techniques after specified target policy updates (i.e., training iteration) for different services. At first, the training dataset is selected as $L \in \{5,10,20,25,30,35,40,45,50\}$, while for the evaluation $L=15$. Next, we consider the training dataset as $L \in \{5,10,15,20,25,30,35,45,50\}$ while for the evaluation $L=40$. Hence, in both cases, the evaluation dataset is different from the training dataset, which helps understand the performance of different techniques when receiving new service offloading requests.
\par
Fig~\ref{fig:ExecutionTimeIteration} represents the average execution time of all techniques in this study. As the number of iterations increases, the execution time of all techniques decreases, meaning that target policy updates lead to better service offloading decisions. However, the convergence speed of service offloading techniques is clearly different. $\mu$-DDRL converges faster to the optimal solution compared to MRLCO, DRLCO, and Double DQN in both cases, as shown in Fig~\ref{fig:ExecutionTimeIteration_15microservices} and Fig~\ref{fig:ExecutionTimeIteration_40microservices}. The V-trace and clipping techniques used in $\mu$-DDRL help for more efficient training on experience trajectories, leading to faster convergence to optimal service offloading solutions. Besides, the learner in $\mu$-DDRL gets the benefit of more diverse experience trajectories in each training iteration on the learner compared to other techniques because it employs experience trajectories generated by several distributed actors. Also, comparing Fig.~ Fig~\ref{fig:ExecutionTimeIteration_15microservices} and Fig~\ref{fig:ExecutionTimeIteration_40microservices} demonstrates that the complexity of services has a direct impact on the performance of service offloading techniques, where suitable service offloading solutions for more complex services require more experience trajectories and policy updates. In addition, DRLCO reaches better service offloading solutions compared to Double DQN and MRLCO but its convergence speed is slower than $\mu$-DDRL. Accordingly, the superiority of $\mu$-DDRL is obtaining the best solution and also converging faster towards best solution.
\begin{figure*}[!t]
    \centering 
    \includegraphics[height=4cm, width=\linewidth]{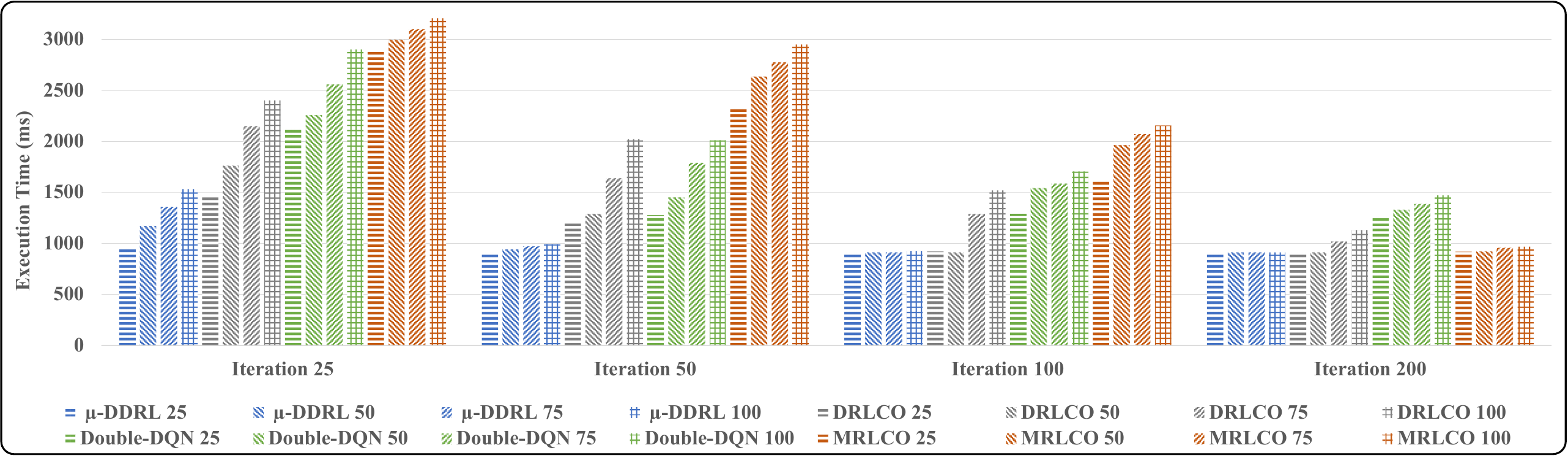}
    \caption{System size analysis}
    \label{fig:SystemSizeAnalysis}
\end{figure*} 
\subsubsection{System Size Analysis}
This experiment represents the performance of service offloading techniques when the number of surrogate computing servers increases. The larger number of servers leads to increased search space and the number of features that directly affect the complexity of the service offloading problem. In this experiment, we consider a different number of servers, where $M \in {25,50,75,100}$. Moreover, we consider the training dataset as $L \in \{5,10,15,20,25,30,35,45,50\}$, while for the evaluation, we use $L=40$. 
\par
Fig.~\ref{fig:SystemSizeAnalysis} presents the obtained execution time of all service offloading techniques while considering different number of servers after different target policy updates (i.e., training iterations). The execution time obtained from each technique decreases as the number of training iterations increases, which exactly follows the pattern shown in the previous experiment. Moreover, as the number of servers increases, the performance of service offloading techniques within the same iteration number is diminished. The background reason is that when the search space and the number of features increase, the DRL-based service offloading techniques require more interactions with the environment, diverse experience trajectories, and training to precisely obtain the offloading solutions. Although the pattern is the same for all techniques, $\mu$-DDRL converges faster to more suitable service offloading solutions and outperforms other techniques even when the system size grows. This is mainly because $\mu$-DDRL uses the shared experience trajectories (i.e. steps in the environment) of different actors in each training iteration, which are usually more diverse compared to experience trajectories of non-DDRL counterparts. Moreover, $\mu$-DDRL training is more efficient than DRLCO because it works on actual experience trajectories forwarded from different actors, while the main policy in DRLCO is trained based on the parameters forwarded from different actors. Thus, $\mu$-DDRL offers higher scalability and adaptability features.
\subsubsection{Speedup Analysis}
The speedup analysis studies how long it takes for DRL agents to obtain a predefined number of experience trajectories. The faster interactions with the computing environment, the more experience trajectories, and the higher potential for faster training. Similarly to \cite{tuli2020dynamic,goudarzi2021distributedDDRL}, we define the speedup as follows:
\begin{equation}
    SP=\frac{Time_R}{Time_T}
\end{equation}
where $Time_R$ is the reference time defined as the required time that the APPO technique with 1 worker reaches 150,000 sample steps in the environment. Also, $Time_T$ is the technique time within which each technique reaches the specified sample steps.
\par
\begin{figure}[!t]
    \centering 
    \includegraphics[height=3.5cm, width=\linewidth]{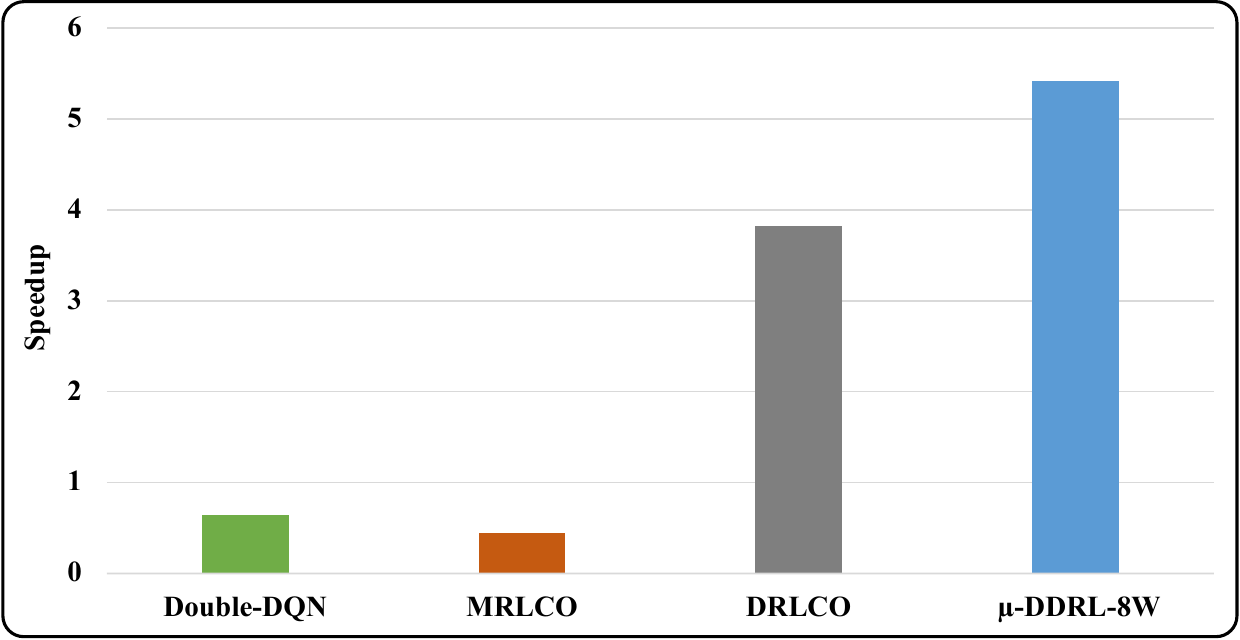}
    \caption{Speedup Analysis}
    \label{fig:Speedup}
\end{figure}

Fig~\ref{fig:Speedup} presents the speedup results of $\mu$-DDRL with 8 workers, MRLCO, DRLCO, and Double DQN techniques. The results depict that $\mu$-DDRL works roughly 8 to 12 times faster than Double DQN and MRLCO techniques in terms of $SP$. Moreover, the speedup of $\mu$-DDRL is about 40\% higher than the DRLCO technique. Therefore, the learning time of the $\mu$-DDRL agent is significantly lower than other techniques, which helps reduce the startup time of DRL agents (i.e., the time DRL agents require to converge) and also provides greater adaptability to highly dynamic and stochastic computing environments, such as fog computing. 
\subsubsection{Decision Time Overhead Analysis}
\begin{figure}[!t]
    \centering 
    \includegraphics[height=3.5cm, width=\linewidth]{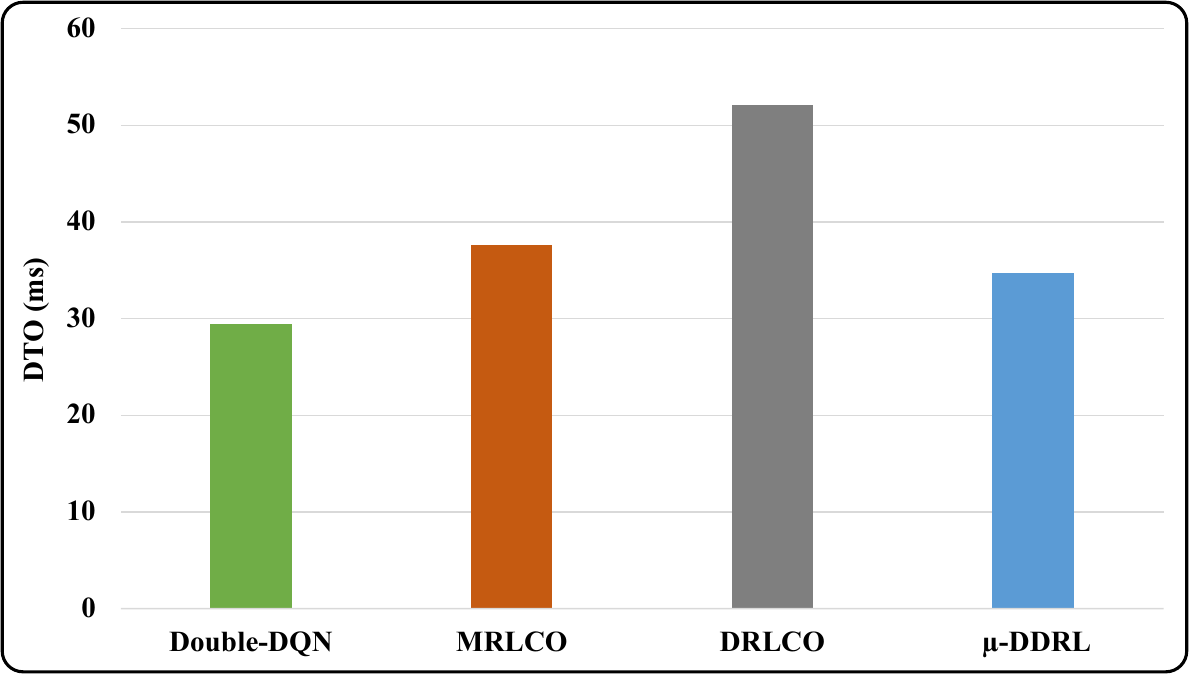}
    \caption{Decision Time Overhead Analysis}
    \label{fig:DTO}
\end{figure}

In this experiment, we study the average Decision Time Overhead (DTO) of all service offloading techniques. DTO represents the average required amount of time that techniques require to make an offload decision for each service with several tasks. For evaluation, we consider services that contain 40 tasks (i.e., $L=40$). As Fig.~\ref{fig:DTO} depicts, while $\mu$-DDRL outperforms MRLCO and DRLCO, it has a higher DTO compared to Double DQN technique. However, because $\mu$-DDRL converges faster to more suitable solutions and its higher efficiency in the training and experience trajectory generation, the negligible increase of DTO in worst-case scenarios is acceptable.
\subsubsection{Optimality Analysis}
In this experiment, we are evaluating the effectiveness of our proposed solution by comparing its performance against the optimal results. To achieve the optimal results, we employed MiniZinc\footnote{https://www.minizinc.org/}, which allows the integration of various optimization solvers, to explore all possible candidate configurations for service offloading. Due to the extensive time required to find the optimal solution, in this experiment, we reduce the number of candidate servers to 8 (i.e., $|\mathcal{M}|=8$) and the number of tasks for the server to 10 (i.e., $L=10$) to reduce the search space of the problem.
\par
\begin{figure}[!t]
    \centering 
    \includegraphics[height=3.5cm, width=\linewidth]{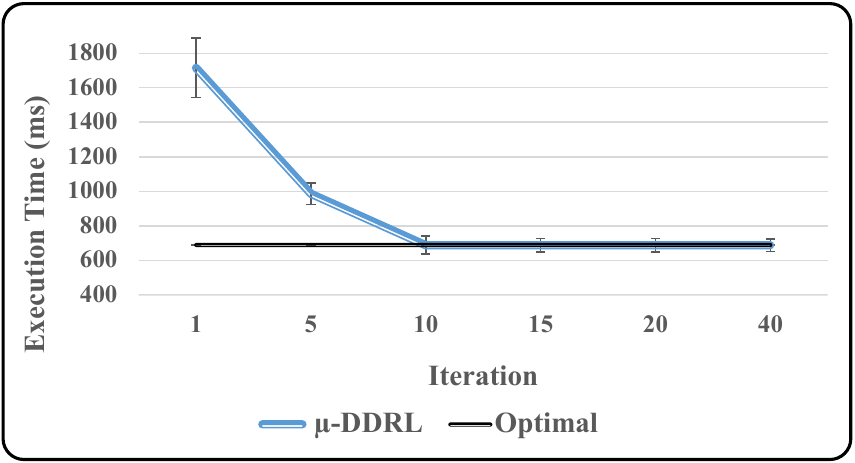}
    \caption{Optimality Analysis}
    \label{fig:optimal}
\end{figure}

Fig.~\ref{fig:optimal} shows the performance of $\mu$-DDRL compared to the optimal results, obtained from Minizinc, in terms of execution time. The results illustrate that $\mu$ -DDRL can reach the optimal solution after 10 training iterations. As the search space for the problem grows, $\mu$-DDRL requires extra exploration and training to converge to optimal solutions.
\section{Conclusions and Future Work}
\label{conclusion}
In this paper, we proposed a distributed DRL technique, called $\mu$-DDRL, for efficient DAG-based service offloading for complex IoT services in highly stochastic and dynamic fog computing environments. We designed an MDP model for service offloading and defined a reward function to minimize the execution time of services while meeting their deadlines. Next, we proposed $\mu$-DDRL as a service offloading technique working based on actor-critic architecture and Asynchronous Proximal Policy Optimization (APPO). The designed DDRL framework enables multiple parallel actors (i.e., brokers that make service offloading decisions) to work simultaneously in a computing environment and share their experience trajectories with the learner for more efficient and accurate training of the target policy. Moreover, we used two off-policy correction techniques, called PPO Clipping and V-trace, to further improve the convergence speed of $\mu$-DDRL towards the optimal service offloading solution. Based on extensive experiments, we demonstrate that $\mu$ DDRL outperforms its counterparts in terms of scalability and adaptability to highly dynamic and stochastic computing environments and different service models. Furthermore, experiments show that $\mu$-DDRL can obtain service offloading solutions to reduce the execution of IoT services by up to 60\% compared to its counterparts. 
\par  
As part of future work, we plan to consider DDRL techniques for IoT services with dynamic mobility scenarios. Moreover, we plan to update the reward function to enable dynamic service migration among different surrogate servers.

\bibliographystyle{IEEEtran}
\bibliography{ref}

\newpage
\appendices
\section{Parameters and respective Definitions}
\label{appendix:paramtersandnotation}
\begin{table*}[!h]
\vspace{-15cm}
\caption{Parameters and respective Definitions}
\label{tab:parametersandnotation}
\resizebox{1\textwidth}{!}{%
\renewcommand{\arraystretch}{1.2}
\begin{tabular}{|c|l|c|l|}
\hline
Parameter & Definition & Parameter & Definition \\ \hline
   $G$       &   A service represented as a DAG.         &     $\mathcal{V}$      & The set of tasks within each service $G$.           \\ \hline
   $v_i$       & The $i$th task in a set of tasks $\mathcal{V}$.           & $\mathcal{E}$          & \begin{tabular}[c]{@{}l@{}}The set of edges denoting the data dependencies between tasks \\ for each service $G$. \end{tabular}            \\ \hline
    $e_{i,j}$      &  A data flow between $v_i$ (parent) and $v_j$ (child)           & $L$          & The maximum number of tasks in a service $G$           \\ \hline
    $e_{i,j}^{w}$       &   The amount of input data that task $v_{j}$ receives from task $v_i$.          &  $v_{j}^{w}$         & The number of CPU cycles required for the processing of the task.            \\ \hline
    $v_{j}^{ram}$     &  The required amount of RAM required to run the task.       &  $\zeta_{v_j}$    &The maximum tolerable delay for the task.           \\ \hline
    $\mathcal{P}(v_{j})$       & The set of all predecessor tasks of $v_j$.           &  $\mathcal{M}$         &   The set of available servers         \\ \hline
    $m^{y,z}$       & \begin{tabular}[c]{@{}l@{}}A server in which $y$ illustrates the server's type and $z$ presents\\ the index of the corresponding server's type. \end{tabular}            & $x_{v_j}$          &  The offloading configuration of task $v_j$.          \\ \hline
     $\mathcal{X}$      &  The offloading configuration set of a service.          & $CP(v_i)$           &   \begin{tabular}[c]{@{}l@{}} A function to indicate whether or not each task is on the critical\\  path of the service. \end{tabular}         \\ \hline
     $\mathcal{CP}$      & The set of tasks on the critical path of the service.           & $\mathscr{T}_{x_{v_j}}$          &   The execution time of each task $v_j$.         \\ \hline
     $\mathscr{T}_{x_{v_j}}^{proc}$      &  The processing time of the task on the corresponding server.          & $\mathscr{T}_{x_{v_j}}^{input}$      &   The time it takes for the input data to become available for that task.         \\ \hline
     $f^{s}_{x_{v_j}}$      &  The processing speed of the corresponding assigned server.          & $b(x_{v_i},x_{v_j})$           &  \begin{tabular}[c]{@{}l@{}} The data rate (i.e., bandwidth) between the selected servers for the\\ execution of $v_i$ and $v_j$.\end{tabular}         \\ \hline
     $l(x_{v_i},x_{v_j})$       &  The communication latency between two servers.          &  $\eth^c$          & The propagation speed for the communication medium.          \\ \hline
     $d(x_{v_i},x_{v_j})$      & \begin{tabular}[c]{@{}l@{}}The Euclidean distance between the coordinates of the participating\\ servers in the Cartesian coordinate system. \end{tabular}            &  $SS(x_{v_i},x_{v_j})$         & \begin{tabular}[c]{@{}l@{}} A function that indicates whether the servers assigned to each pair of\\ tasks are the same (that is, $0$ if $x_{v_i} = x_{v_j}$) or different (i.e., $1$ if $x_{v_i} \neq x_{v_j}$). \end{tabular}           \\ \hline
     $\mathcal{T}(\mathcal{X})$      & The execution time of each service based on its configuration.          & $IA(v_j,m^{y,z})$          &  \begin{tabular}[c]{@{}l@{}}An indicator function to check if the task $v_j$ is assigned to the server\\ $m^{y,z}$ ($IA=1$) or not ($IA=0$).\end{tabular}.          \\ \hline
     $\mathbb{S}$       & The state space in MDP.          &  $\mathbb{A}$        &  The action space in MDP.          \\ \hline
     $\mathbb{P}$      & A state transition between states in MDP.           &  $\mathbb{R}$         & The reward function in MDP.          \\ \hline
     $\gamma$      & The discount factor in MDP.           & $\pi(a_t|s_t)$          & The agent policy.           \\ \hline
     $s_t$      &  The state of environment at time step $t$.          & $a_t$          & The action at time step $t$.           \\ \hline

     $r_t$& The reward at time step $t$         &  $F^{\mathcal{M}}_{t}$         & The feature vector of all $M$ servers at time step $t$.\\ \hline
     $F^{v_j}_{t}$ & The feature vectors of the current task in a service.         &  $f_i^{m^{y,z}}$          &The $i$th feature corresponding to the server $m^{y,z}$. \\ \hline
     $f_i^{v_j}$ & The $i$th feature of the task $v_j$.         &           $\Phi$ & The failure penalty to penalize actions that violate $\zeta_{v_j}$.\\ \hline 
       
\end{tabular}%
}
\end{table*}
A summary of the parameters and their respective definitions is presented in table~\ref{tab:parametersandnotation}.

\end{document}